\documentclass[twocolumn,floatfix, nofootinbib,prd,preprintnumbers,showpacs,showkeys,superscriptaddress,longbibliography]{revtex4-1} 

\usepackage{amssymb}
\usepackage[mathscr]{euscript}
\usepackage{amsmath}
\usepackage{graphicx}
\usepackage{dcolumn}
\usepackage{bm}
\usepackage{epsfig}
\usepackage{amssymb,latexsym,mathrsfs}
\usepackage{graphicx}
\usepackage{color}
\usepackage{hyperref}
\usepackage{float}
\usepackage[thinlines]{easytable}
\usepackage{multirow}
\usepackage{diagbox}
\usepackage{varwidth}
\usepackage{array}
\newcolumntype{P}[1]{>{\centering\arraybackslash}p{#1}}

\newcolumntype{M}[1]{>{\centering\arraybackslash}m{#1}}
\newcolumntype{N}{@{}m{0pt}@{}}

\usepackage{amsmath}
\newcommand*\diff{\mathop{}\!\mathrm{d}}

\usepackage{amssymb}
\usepackage{pifont}
\hypersetup{
    colorlinks=true,
    linkcolor=red,
    citecolor=blue,
}

\usepackage{amsmath}
\usepackage{amssymb}
\usepackage[caption=false]{subfig}
\usepackage{hyperref}
\usepackage{url}
\usepackage{xcolor}
\usepackage{color}
\definecolor{amaranth}{rgb}{0.9, 0.17, 0.31}
\definecolor{purple(munsell)}{rgb}{0.62, 0.0, 0.77}
\definecolor{americanrose}{rgb}{1.0, 0.01, 0.24}
\definecolor{palatinateblue}{rgb}{0.15, 0.23, 0.89}
\definecolor{royalblue(web)}{rgb}{0.25, 0.41, 0.88}
\definecolor{hanpurple}{rgb}{0.32, 0.09, 0.98}
\definecolor{beaublue}{rgb}{0.74, 0.83, 0.9}
\definecolor{carminered}{rgb}{1.0, 0.0, 0.22}
\definecolor{brightpink}{rgb}{1.0, 0.0, 0.5}
\definecolor{vividviolet}{rgb}{0.62, 0.0, 1.0}

\definecolor{electron}{rgb}{1.0, 0.67, 0.22}
\hypersetup{ linktoc=all,
    colorlinks, linkcolor={palatinateblue},
    citecolor={brightpink}, urlcolor={amaranth}}

\newcommand{\be}{\begin{equation}}
\newcommand{\ee}{\end{equation}}
\newcommand{\bs}{\begin{split}} 
\newcommand{\bea}{\begin{eqnarray}}
\newcommand{\eea}{\end{eqnarray}}

\newcommand{\ievlev}[1]{\textcolor{magenta}{[{\bf EI}: #1]}}
\newcommand{\kn}[1]{\textcolor{blue}{[{\bf KN}: #1]}} 

\newcommand{\bes}{\begin{subequations}}
\newcommand{\ees}{\end{subequations}}

\newcommand{\bo}{\raise-1mm\hbox{\Large$\Box$}}

\usepackage{physics}
\usepackage{csquotes}
\usepackage{mathtools}

\begin{document}

\preprint{FTPI-MINN-24-07}

\title{
   Classical acceleration temperature from evaporated black hole remnants and accelerated electron-mirror radiation
    }

\author{Kuan-Nan Lin}
\affiliation{Department of Physics and Center for Theoretical Sciences,\\ National Taiwan University, Taipei 10617, Taiwan.}
\affiliation{Leung Center for Cosmology and Particle Astrophysics,\\
National Taiwan University, Taipei 10617, Taiwan.}

\author{Evgenii Ievlev}
\email{ievle001@umn.edu}
\affiliation{William I. Fine Theoretical Physics Institute, School of Physics and Astronomy,\\
University of Minnesota, Minneapolis, MN 55455, USA.}
\author{Michael R.R. Good}
\email{michael.good@nu.edu.kz}
\affiliation{Physics Department \& Energetic Cosmos Laboratory,\\ Nazarbayev University,
Astana 010000, Qazaqstan.}
\affiliation{Leung Center for Cosmology and Particle Astrophysics,\\
National Taiwan University, Taipei 10617, Taiwan.}
\author{Pisin Chen}
\email{pisinchen@phys.ntu.edu.tw}
\affiliation{Department of Physics and Center for Theoretical Sciences,\\ National Taiwan University, Taipei 10617, Taiwan.}
\affiliation{Leung Center for Cosmology and Particle Astrophysics,\\
National Taiwan University, Taipei 10617, Taiwan.}
\affiliation{Kavli Institute for Particle Astrophysics and Cosmology, SLAC,\\ Stanford University, Stanford, CA 94305, USA.}

\begin{abstract} 
We investigate the radiation from accelerating electrons with asymptotic constant velocity and their analog signatures as evaporating black holes with left-over remnants. 
We find high-speed electrons, while having a high temperature, correspond to low-temperature analog remnants. 
\end{abstract} 

\date{\today} 

\maketitle

\section{Introduction}

The intricate interplay between uniform acceleration and temperature has been extensively explored within the framework of quantum scalar fields, notably through the well-established Unruh effect \cite{Fulling:1972md,Davies:1974th,unruh76}. However, a classical understanding of the link between non-uniform acceleration and temperature (e.g. from a Planck electromagnetic radiation spectrum) has remained largely unexplored until recent efforts \cite{Good:2022eub,Ievlev:2023inj,Good:2022xin,Ievlev:2023bzk,IEVLEV2023129131}. This gap in the existing literature calls for a significant shift in focus from the quantum-only approach to an investigation of classical acceleration temperature (CAT).  

A moving point charge along non-uniform acceleration results in CAT \cite{Lynch:2022rqx}.  This nascent study of the classical relationship between acceleration and temperature presents an intriguing avenue for research, promising to unveil new insights into thermal dynamics. By studying CAT, we aim to help bridge this gap and expand our understanding of the fundamental interplay between motion and thermal properties within experimentally accessible, classically tractable systems, e.g. thermal beta decay \cite{Lynch:2022rqx}.  In particular, this involves analyzing speed-dependent thermal radiation spectra from a non-uniformly accelerated moving point charge.

Our approach will exploit the important concept of analog systems that have, with the growing complexity of theoretical models, proven fruitful for drawing fresh insights from both quantum and classical perspectives.
This paper considers such a classical-quantum analog between moving mirrors, black hole remnants, and accelerated electrons.

In 1992 Wilczek \cite{wilczek1993quantum} noticed that moving mirrors \cite{DeWitt:1975ys,Davies:1976hi,Davies:1977yv} are analogs for black hole remnants \cite{Chen:2014jwq}. Specifically, asymptotic constant subliminal speed velocity trajectories suffer no information loss (quantum purity) \cite{Good_2015BirthCry}.  These particular trajectories are not a model for complete black hole evaporation \cite{Good:2019tnf}, but instead, their end-state signifies a remnant that highly red-shifts frequencies at late times \cite{Good:2016atu}.

Unruh and Wald in 1982 \cite{Unruh:1982ic}, and Nikishov and Ritus in 1995 \cite{Nikishov:1995qs}, noticed that radiation from accelerated electrons and moving mirrors are identical\footnote{Up to a factor of $4\pi \alpha_{\textrm{fs}}$, where $\alpha_{\textrm{fs}}$ is the fine structure constant.}.   
Further development on this connection has substantiated the exact correspondence, see e.g. \cite{Ievlev:2023inj,Ievlev:2023bzk,Good:2022eub,Ievlev:2023inj,Good:2022xin,Lynch:2022rqx,Ritus:1999eu,Ritus:2002rq,Ritus:2003wu,Zhakenuly:2021pfm,Ritus:2022bph}.

This work examines the CAT radiation emitted by accelerating electrons with a constant velocity at infinity to help understand black hole evaporation leading to the emergence of a remnant. To accomplish this goal, we focus on three worldlines that are of particular geometric significance, namely the remnant versions of the Schwarzschild \cite{Schwarzschild:1916uq}, CGHS \cite{Callan:1992rs} and BTZ black hole \cite{Banados:1992wn}.

We limit the treatment to the Schwarzschild, CGHS, and BTZ trajectories.  Other interesting future-drift motions asymptotically approach a constant speed of light. 
 These have commensurate asymptotic approach to zero proper acceleration and are the so-called `inertial-null' trajectories; e.g. Proex \cite{good2013time,Good:2016yht} (which is the light speed case of the beta decay trajectory \cite{Good:2022eub,Ievlev:2023inj,Good:2022xin,Lynch:2022rqx}), an inertial horizon \cite{Good:2020uff}, and Light \& Airy \cite{Good:2021dkh,Good:2022gvk}. However, these (and the future drift GO trajectory \cite{Good:2023hsv} and Drift-Davies-Fulling \cite{Good:2017kjr}) 
will be left outside this study's scope because they do not correspond to well-known\footnote{Extremal black holes are also excluded because their analogs approach non-zero albeit uniform proper acceleration \cite{Liberati_2000, good2020extreme}} black holes. 

The paper is organized as follows. 
In Sec.~\ref{sec:duality_intro} we provide some background on the correspondence between moving mirrors, accelerated electrons, and general relativity.
Next, we treat three examples, focusing on the accelerated electrons corresponding to: 
Schwarzschild remnant in Sec.~\ref{sec:SR_trajectory},
CGHS remnant in Sec.~\ref{sec:CGHS_trajectory},
and BTZ system in Sec.~\ref{sec:BTZ}.
The results are summarized with the conclusions in Sec.~\ref{sec:conclusions}. We use units $\hbar = \mu_0 = \epsilon_0 = c = k_B = G = 1$, and $e^2 = 4\pi \alpha_{\textrm{fs}} \approx 0.0917$.

\section{Elements of the electron-mirror-gravity triality}
\label{sec:duality_intro}

\subsection{Accelerating electrons and mirrors}

This section reviews the duality between the radiation from an electron in 3+1d and a moving mirror in 1+1d.
Some of the relevant results are relatively recent.

The moving mirror, or dynamical Casimir effect, occurs when considering a quantum field with an accelerated boundary. This means the following.
Consider a 1+1d space $(t,z)$. 
Take a quantum massless scalar field $\Phi$ with a Dirichlet boundary condition: $\Phi=0$ at $z = z(t)$ with a pre-set function $z(t)$.
Then, if the boundary $z_m(t)$ moves with acceleration, there will be an observable energy flux.
One can calculate the beta Bogolubov coefficients\footnote{Subscript $R$ refers to the right side of the mirror. The left side can also be considered in the same way.}
$\beta^R_{pq}$ responsible for mixing of creation-annihilation operators in this dynamical setup \cite{DeWitt:1975ys,Davies:1976hi,Davies:1977yv}.
Here, $p$ and $q$ are the frequencies of the corresponding field modes.

Now consider an electron in 3+1d moving in a straight line along the $z$-axis with some (externally forced) acceleration. 
According to classical electrodynamics, such an electron would radiate electromagnetic waves, and one can calculate the corresponding spectral distribution $\diff I / \diff \Omega$, where $\Omega$ is the solid angle, and $I$ is the intensity (the power per unit frequency).

As it turns out, if one takes the mirror and the electron set on the same trajectory $z = z(t)$, the radiation spectra in these two seemingly different problems are related.
The exact functional correspondence was recently derived in
\cite{Ievlev:2023bzk,Ievlev:2023inj}:
\begin{equation}
\begin{aligned}
	&\frac{\diff{I}}{\diff{\Omega}}(\omega,\cos\theta) = \frac{e^2 \omega^2}{4\pi} |\beta^R_{pq}|^2 \\
	&p + q = \omega \,, \quad p - q = \omega \cos\theta \\
	&p = \omega \frac{1 + \cos\theta }{2} \,, \quad 	q = \omega \frac{1 - \cos\theta }{2}
\end{aligned}
\label{recipe_dIdOmega_from_mirror}
\end{equation}
Here, $p$ and $q$ are the frequencies in the mirror setup.
We use this notation to avoid confusing it with the 3+1 photon frequency $\omega$. The physical context of the election-mirror duality for the mirror needs to account for both sides of the mirror. This ensures an exact correspondence to the radiation emitted by an electron.  However, conveniently, the recipe Eq.~(\ref{recipe_dIdOmega_from_mirror}), due to the parity symmetry $\beta^R_{pq} = \beta^L_{qp}$ needs only the right-side beta Bogoliubov coefficient, $\beta^R_{pq}$.   
Table~\ref{tab:limits} compares some limiting cases on both sides of this correspondence.

\begin{table}[ht] 
\begin{tabular}{l | l }
Mirror & \ Electron   \\ \hline
\rule{0pt}{1.05\normalbaselineskip}moves to the left\ \ & \ moves down $z \to - \infty$  \\
	high-freq.\ $q \gg p$ & \ blueshift-forward $\theta \to \pi$, $q \approx \omega$ \\
	low-freq. $q \ll p$ & \ redshift-recede $\theta \to 0$, $p \approx \omega$
\end{tabular} \\ 
\caption{Correspondence between mirror and electron properties in various limits \cite{Ievlev:2023inj}.} 
\label{tab:limits} 
\end{table}

\subsection{Black hole correspondence}

As discussed in the Introduction, moving mirrors have been studied in connection with black hole radiation.
For many collapsing geometries, it is possible to construct corresponding coefficients that characterize the spectrum.  The radiation spectra (or the Bogolubov coefficients) are equal on both sides of this correspondence.
It is natural to ask how a gravitational system (black hole) can be mimicked by a non-gravitational system (mirror or electron). 
This can explained by noticing the scales and parameters involved in each system.

In the semi-classical mirror picture, the mirror-induced radiation (MIR) is quantum-originated, and the relevant constants are the Planck constant $\hbar$ and an acceleration parameter $\kappa$.
The mirror's trajectory is exactly known and, therefore, `classical' (the field, however, is quantized).  The acceleration scale $\kappa_\text{mir}$ is a free parameter, and one can always stretch the trajectory to achieve a desired scale.

In the semi-classical black hole picture, the essential constants involved are the Planck constant $\hbar$ and the surface gravity $\kappa_\text{grav}$, as Hawking radiation is considered both a quantum and a gravitational phenomenon.
In terms of the black hole mass $M$ (which is a free parameter), we have $\kappa_\text{grav} = 1/(4 G M)$ (taking the speed of light $c=1$).

In the classical electrodynamics picture, the relevant constants when considering radiation from an accelerated electron are the electric charge $e$ and the acceleration parameter $\kappa_\text{el}$.
One can build a parameter $\mu_0 c e^2$, which in SI units has the same dimensionality as $\hbar$ (see the discussion in \cite{Good:2022xin}).

Thus, in each of the three systems, we have a fixed\footnote{It could be argued that the fine structure constant $\alpha_{\textrm{fs}}$ is subject to the RG flow and thus not exactly fixed. However, this energy dependence is very weak: between 1 MeV and 100 GeV it changes only by about 6 percent. In any case, at energy scales where $\alpha_{\textrm{fs}}$ is of order 1, the approximation of classical electrodynamics has long ceased to be applicable.}
parameter with the dimension of the action, $\hbar$ or $\mu_0 c e^2$,
and a free parameter with the dimension of acceleration, $\kappa_\text{mir}$, $\kappa_\text{grav}$ or $\kappa_\text{el}$.
In this triality, the three latter parameters are identified with each other; below, we write $\kappa$ without a subscript.
The acceleration parameter of the flat-space moving mirror and accelerated electron plays the role of the gravitational constant in the curved spacetime.

In the black hole picture, Hawking temperature is given by $T_{H}=\hbar\kappa/(2\pi)=10^{-6}M_{\odot}/M$ Kelvin, where $\kappa=1/(4GM)$ is the black hole surface gravity. The temperature is too low for astrophysical black hole detections. However, in the mirror picture, $\kappa$ is a free parameter, so one can, in principle, detect exact analog Hawking radiation of the motion-induced radiation (MIR) much easier than black hole radiation by tuning the value of $\kappa$. 

The caption in Table \ref{role_of_parameters} summarizes the role of the parameters in the black hole-mirror-electron correspondence (see also \cite{Good:2022xin} for a detailed discussion). The role can be considered a reduction process from the black hole to the mirror and then to the electron picture. Interestingly, upon each reduction, the difficulty of experimental concerns significantly reduces.

\begin{table}[ht] 
\centering
%
%
\begin{tabular}{|P{2.0cm}|P{2.0cm}|P{2.0cm}|P{2.0cm}}
Nature & Black Hole & Mirror &  Electron   \\
\hline\hline
Quantum $\hbar$ & Yes & Yes  & $\hbar \to \mu_0 c e^2$ \\[5pt]
Gravity $G$ & Yes & $\dfrac{c^2}{4GM} \to \kappa$ & $\dfrac{c^2}{4GM} \to \kappa$ 
\end{tabular} \\ 
\caption{The role of the parameters in the correspondences depends on whether the system involves quantum fields or curved spacetime (SI units). }
\label{role_of_parameters} 
\end{table}

\subsection{From black holes to remnants}

\begin{figure*}[t]
    \subfloat[Formation of a black hole and its end state as an eternal remnant after evaporation.]{\label{fig:remnant Penrose}\includegraphics[width=0.8\columnwidth]{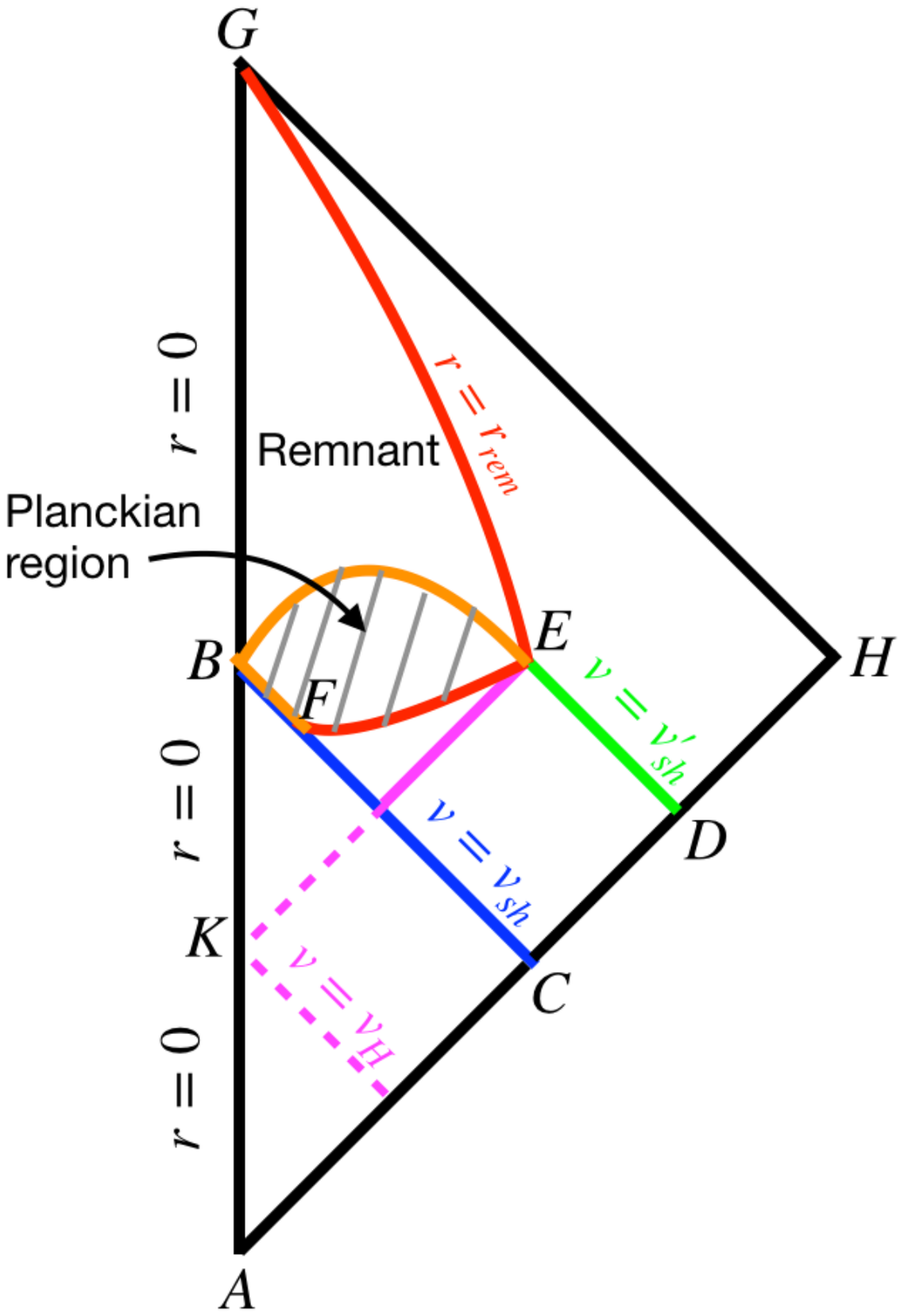}}
    \qquad
    \subfloat[A mirror, with wordline AKBE$'$G, in flat spacetime that has a constant velocity in the future.]{\label{fig:mirror Penrose}\includegraphics[width=0.9\columnwidth]{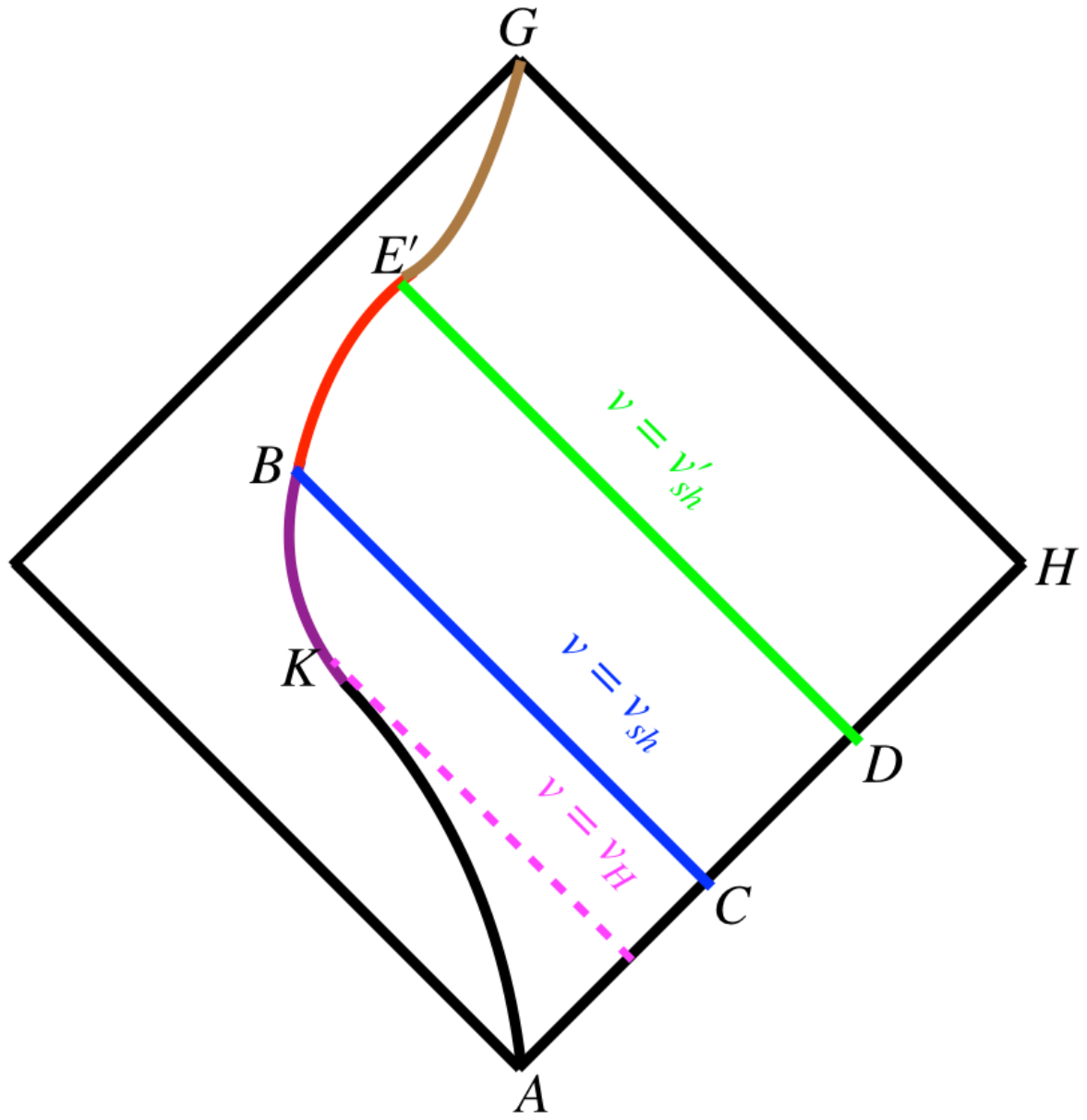}}
\caption{
	The region on the right of the mirror \ref{fig:mirror Penrose} can be mapped into a Penrose diagram that resembles the Penrose diagram of a remnant on the left \ref{fig:remnant Penrose}. In the right figure, segment $AK$ is the period in which the mirror has the typical accelerated trajectory relevant for analog Hawking radiation emission; segments KB and BE$'$ are deceleration phases; segment E$'$G is the phase of constant velocity. The red curve $r=r_{rem}$ in the left figure is the remnant's radius. For further clarity, see similar figures in, e.g., \cite{Good:2018zmx} and fig.2 in \cite{Chen:2014jwq}. BFE is not the apparent horizon.   Basically, (region bounded within) ABFC is the initial flat spacetime. As a null shell carrying positive energy flux collapses along $v=v_{sh}$, a Schwarzschild spacetime CFBFED is formed. Since we consider the remnant as the final state, DHGBE is the final flat spacetime with a (time-like) remnant. BEFB is the region connecting the previous three regions. It is not known exactly what BEFB is. Still, FE is the remnant radius within the horizon (magenta line), and the horizontal line BE, which is not drawn in the figure, will be the singularity if the final state is a black hole instead. Since one doesn't know precisely how the singularity is removed, the shaded, probably Planckian region is inside the remnant radius and thus involves the removal of singularity region BEFB. $v=v_{sh}'$ connects the Schwarzschild and final flat spacetime. It may also be regarded as the evaporation of the original black hole, i.e., the injection of negative energy flux into the original horizon. Since fig.1a and 1b can be related by a coordinate transformation ($r=0$ in fig.1a maps into the mirror trajectory in fig.1b), the null lines $v=v_{sh},v_{sh}'$ in fig.1a will correspond to the null lines shown in fig.1b.}    
	
\label{fig:remnant-mirror Penrose}
\end{figure*}

While the black hole-mirror analogy has been well established for over half a century \cite{Davies:1976hi}, the remnant-mirror duality needs more investigation. Figure \ref{fig:remnant-mirror Penrose} explicitly illustrates the duality. Radiation emitted during the period AK in Figure \ref{fig:mirror Penrose} corresponds to the usual Hawking radiation. The new features in the remnant-mirror duality are as follows.

Radiation emitted during the period KB in Figure \ref{fig:mirror Penrose} corresponds to the emission of a partner particle of the Hawking radiation. Looking at Figure \ref{fig:remnant Penrose}, one notices that the partner particle first undergoes a Schwarzshild region inside the horizon and then passes through a Planckian region within the remnant 
(the terminology is explained in Fig.~\ref{fig:remnant-mirror Penrose}), 
and finally propagates to the future null infinity. In the gravitational system, we do not yet have knowledge about Planck scale physics, so what happens to the partner particle in the Planckian region is unknown. However, looking at Figure \ref{fig:mirror Penrose}, one notices that the mirror motion during KB provides a possible mechanism for better understanding the otherwise unknown Planckian physics. It would be intriguing to develop this duality in more detail.

Radiation emitted during the period BE$'$ in Figure \ref{fig:mirror Penrose} corresponds to field mode emitted from the segment CD falling into the horizon in Figure \ref{fig:remnant Penrose}, infalling modes emitted at CD eventually falls into the supposed horizon (magenta line), passes through the Planckian region, and eventually propagates to the future null infinity. This radiation is neither Hawking radiation nor the partner particles mentioned above, and, interestingly, this emittance may have to do with the well-known existence of negative outgoing energy flux \cite{Walker:1984ya}.  See also \cite{Good:2015nja}, e.g., Fig. 2.  In the case of a (1+1)D perfectly reflecting point mirror, the energy flux $\langle T_{uu} \rangle$ emitted by the mirror to the future null infinity $I^{+}_{R}$ will be negative for some time after the thermal plateau, provided the mirror reaches the future time-like infinity.

Finally, the segment E$'$G in Figure \ref{fig:mirror Penrose} does not radiate since the mirror is inertial, albeit ingoing field modes emitted from the segment DH will undergo significant Doppler redshift upon reflections if the mirror has a high final coasting speed. By passing to the remnant picture \ref{fig:remnant Penrose}, one realizes that, while field modes undergo significant redshift upon leaving the remnant, only those that also pass through the Planckian region can become real particles due to the yet unknown Planck scale physics.

The corresponding Penrose diagrams are identical in the case of a mirror or an electron moving in a flat spacetime. However, a difference can exist in how the radiation is emitted since the two pictures are related by a coordinate transformation of $(p,q)$ and $(\omega,\theta)$ as indicated in Eq.~\eqref{recipe_dIdOmega_from_mirror}. 
For example, analog Hawking radiation in the mirror picture can manifest as exact thermal Larmor radiation  \cite{Ievlev:2023inj} in the election picture.  In addition, while the radiation spectra on both sides of the mirror are generally different, Larmor radiation is symmetric in the azimuthal angular direction. Thus, the radiation emitted by the mirror during different acceleration-deceleration phases, which corresponds to different regimes of $(p,q)$, may neatly be separated into different polar angle $\theta$ regimes for radiation emitted by the electron.

\section{Schwarzschild-remnant} 
\label{sec:SR_trajectory}

In this section, we study an accelerated electron dual to one specific remnant solution. This is interesting because of its asymptotic connection to the radiation spectrum emitted by a Schwarzschild black hole.
Investigated as a 1+1 dimensional moving mirror in \cite{Good:2016atu,Good:2018ell,Myrzakul:2018bhy,Good:2018zmx} this solution was called `the drifting Omex' or `Giant Tortoise mirror' or, `the Schwarzschild-remnant'.  It is a model for finite energy emission during black hole evaporation while having no information loss, whose final state gives rise to a non-trivial Doppler shift to the field modes, indicating the presence of a remnant \cite{wilczek1993quantum}.

\subsection{Trajectory and global properties}

The trajectory equation of motion for the Schwarzschild-remnant system is most simply expressed as a function of space,
\begin{equation}
t(z)=-\frac{z}{s}-\frac{1}{\kappa}e^{2\kappa z/s}.
\label{SR_traj}
\end{equation}
Here $s$ is the final drift speed, and $\kappa$ is the acceleration scale. The independent variable designating lab space coordinate is $z$, while $t$ is coordinate lab time but expressed as the dependent variable. 

Consider an accelerated electron moving along the trajectory Eq.~\eqref{SR_traj}. Then, the total power, $P =e^2\alpha^2/6\pi$ is
\be P = \frac{ 8 e^2 \kappa ^2}{3 \pi}\frac{ s^2 e^{4 \kappa  z/s}}{\left[\left(1+ 2 e^{2 \kappa  z/s}\right)^2-s^2\right]^3}.\label{Schpower}\ee
The power exhibits no known flattening or plateau indicative of extended constant energy emission accompanying thermal equilibrium. As we shall see, despite the explicit Planck factors in the following spectrum, the power $\diff{E}/\diff{u}$ does not indicate a constant emission period. 

\begin{figure}[h]
\includegraphics[width=\columnwidth]{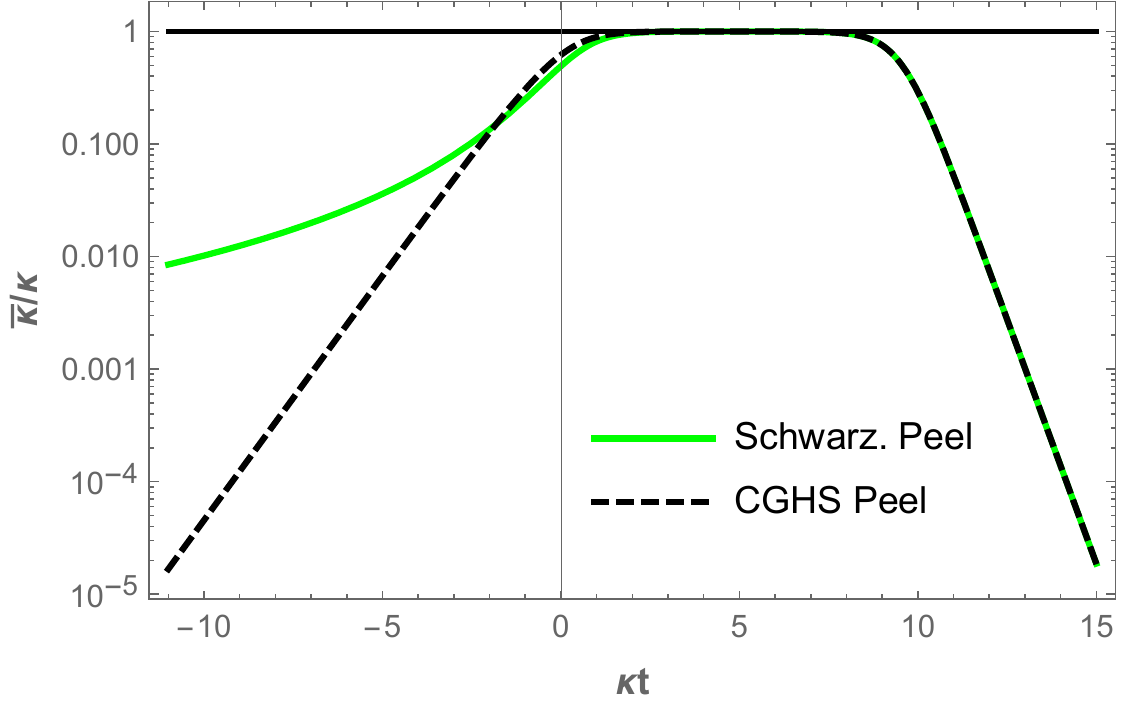} 
 \caption{The Schwarzschild and CGHS peel acceleration, Eq.~(\ref{SR_peel}) and Eq.~(\ref{CGHS_peel}) of their respective electron trajectories, Eq.~(\ref{SR_traj}) and Eq.~(\ref{cghs_traj_deformed}), as functions of dimensionless coordinate time $\kappa t$. Here, the final speed for both is set to the ultra-relativistic final speed 
 $s = 1 - 10^{-8}$ (or rapidity $\eta = 9.55$, $s = \tanh \eta$)
 to demonstrate the flattening plateau which is normalized at magnitude $\bar{\kappa}/\kappa = 1$. Two key takeaways are the future-past asymptotic approach to zero and the intermediate constant peel plateau. 
 }
\label{Fig_Peel}
\end{figure}

However, it is possible to confirm thermality with the dynamics but in a different way.  We can use a thermally relevant acceleration to confirm our intuition that a flattening-type equilibrium accompanies the spectral analysis.  Thermality is confirmed from the peel\footnote{The `peeling function' or effective temperature function, is common in relativistic contexts, see e.g. \cite{Bianchi:2014qua,Barcelo:2010pj,McMaken:2023tft}. Carlitz-Willey pointed out that the thermal moving mirror has a constant peel  \cite{CW2lifetime}.} 
acceleration $\bar{\kappa}$ in the asymptotic limit that $s\to 1$, i.e. ultra-relativistic final drifting speeds. To leading order, in light cone coordinate $u=t-z$,
\be \bar{\kappa} = \frac{\kappa }{\left[W\left(e^{-\kappa u}\right)+1\right]^2}, \label{SR_peel}\ee
which at late times gives
\be \lim_{u\to+\infty}\bar{\kappa} = \kappa,\ee
corroborating thermal Planck-distributed equilibrium below. 
See Figure \ref{Fig_Peel} for a plot of the Schwarzschild peel acceleration as a function of coordinate time $t$ and the associated plateau indicative of thermal emission. 

The energy emitted by the electron is found by integrating the power, Eq.~(\ref{Schpower}), over coordinate time, 
\be E = \frac{e^2\kappa}{24\pi}\left( \left[\gamma^2 -1\right] - \left[\frac{\eta}{s}-1\right]\right).\label{SR_energy0}\ee
Here $\gamma = 1/\sqrt{1-s^2}$ is the Lorentz factor, and $\eta = \tanh^{-1}s$ is the rapidity. The energy is expressed in this form to highlight the appearance of the soft-energy $E_\textrm{BD} \sim (\eta/s -1)$ \cite{Good:2022eub}, distinct from $(\gamma^2-1)$. 

After looking at the dynamics of Eq.~\eqref{SR_traj} and the straightforward calculation of power, peel, and energy, we now understand that the peel (not the power) most readily reveals an indication of thermality. We are ready to use the total finite energy emitted, Eq.~(\ref{SR_energy0}), to confirm the correctness of a spectral analysis. We will now turn to computing the spectrum. 


\subsection{Electromagnetic spectrum \& temperature}

\begin{figure}[h]
\includegraphics[width=0.6\linewidth]{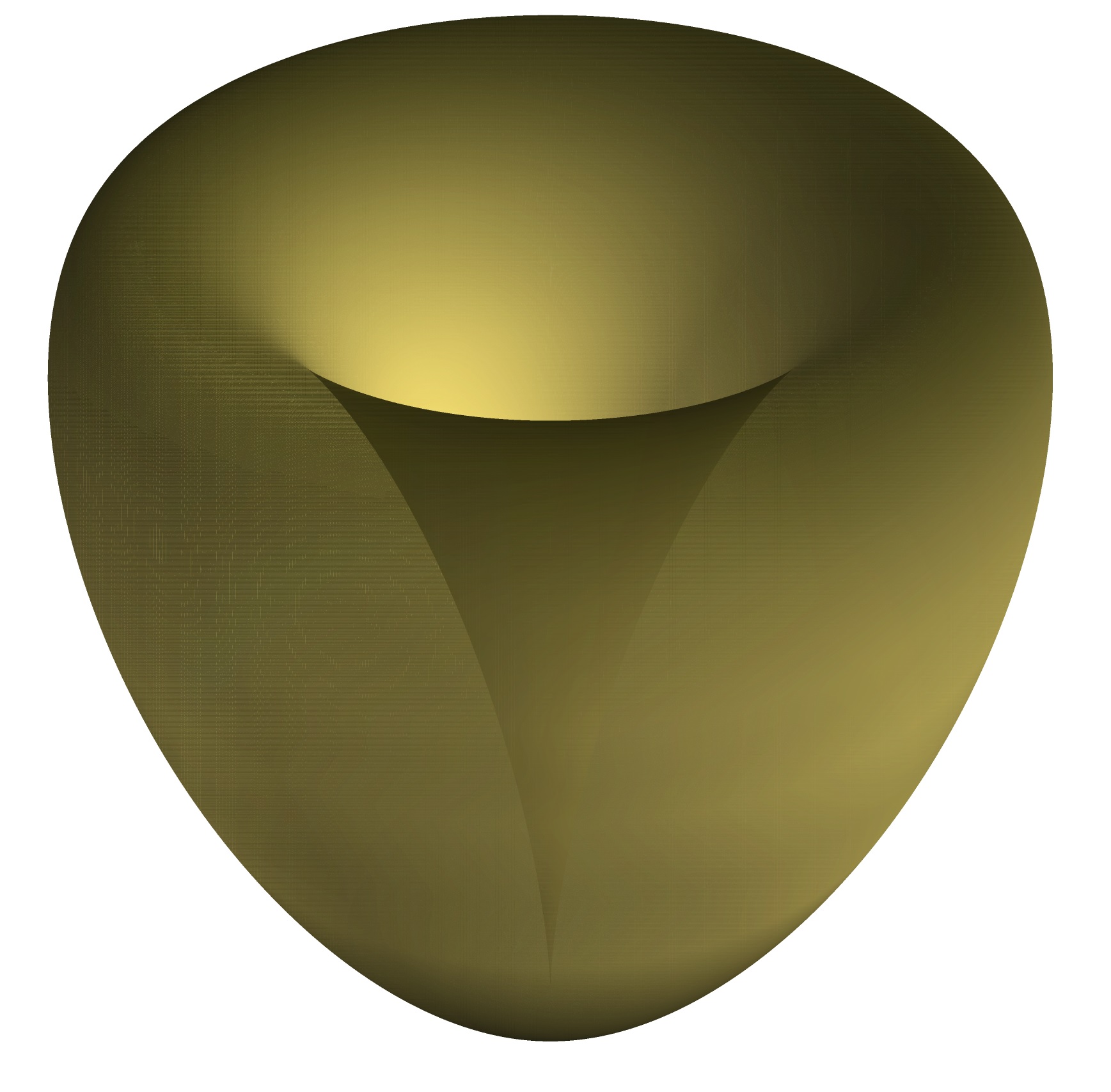} 
 \caption{The Schwarzschild electron spectral distribution, Eq.~(\ref{SR_dis}), with $\kappa = \omega = 1$ and $s=0.9$.
 }
\label{Fig_SR_spec_dis}
\end{figure}

\begin{figure}[h]
\includegraphics[width=0.6\linewidth]{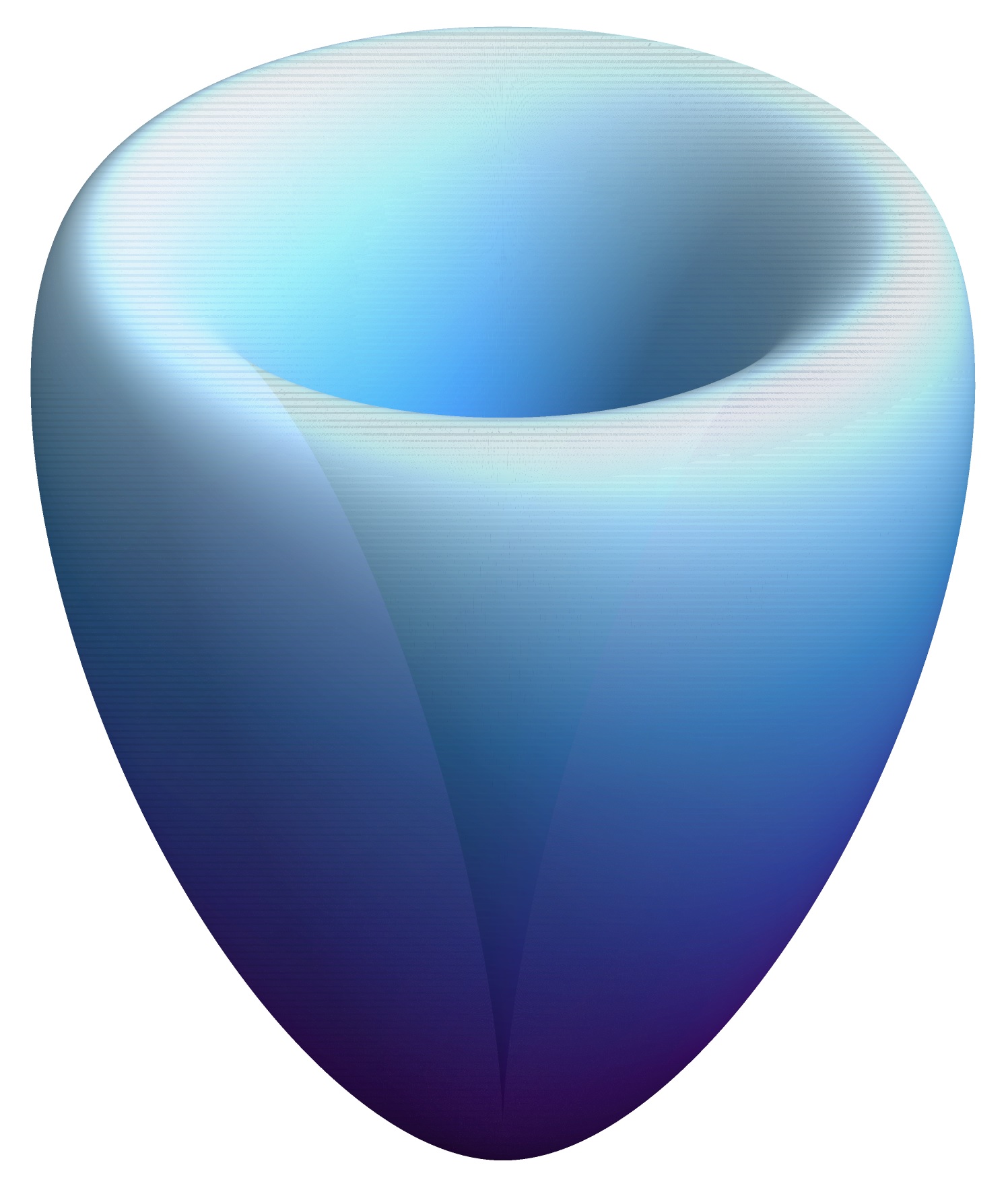} 
 \caption{The Schwarzschild electron energy distribution, Eq.~(\ref{SR_energy_dis}), with $\kappa = 1$ and $s=0.9$.
 }
\label{Fig_SR_energy_dis}
\end{figure}

\begin{figure}[h]
\includegraphics[width=0.4\linewidth]{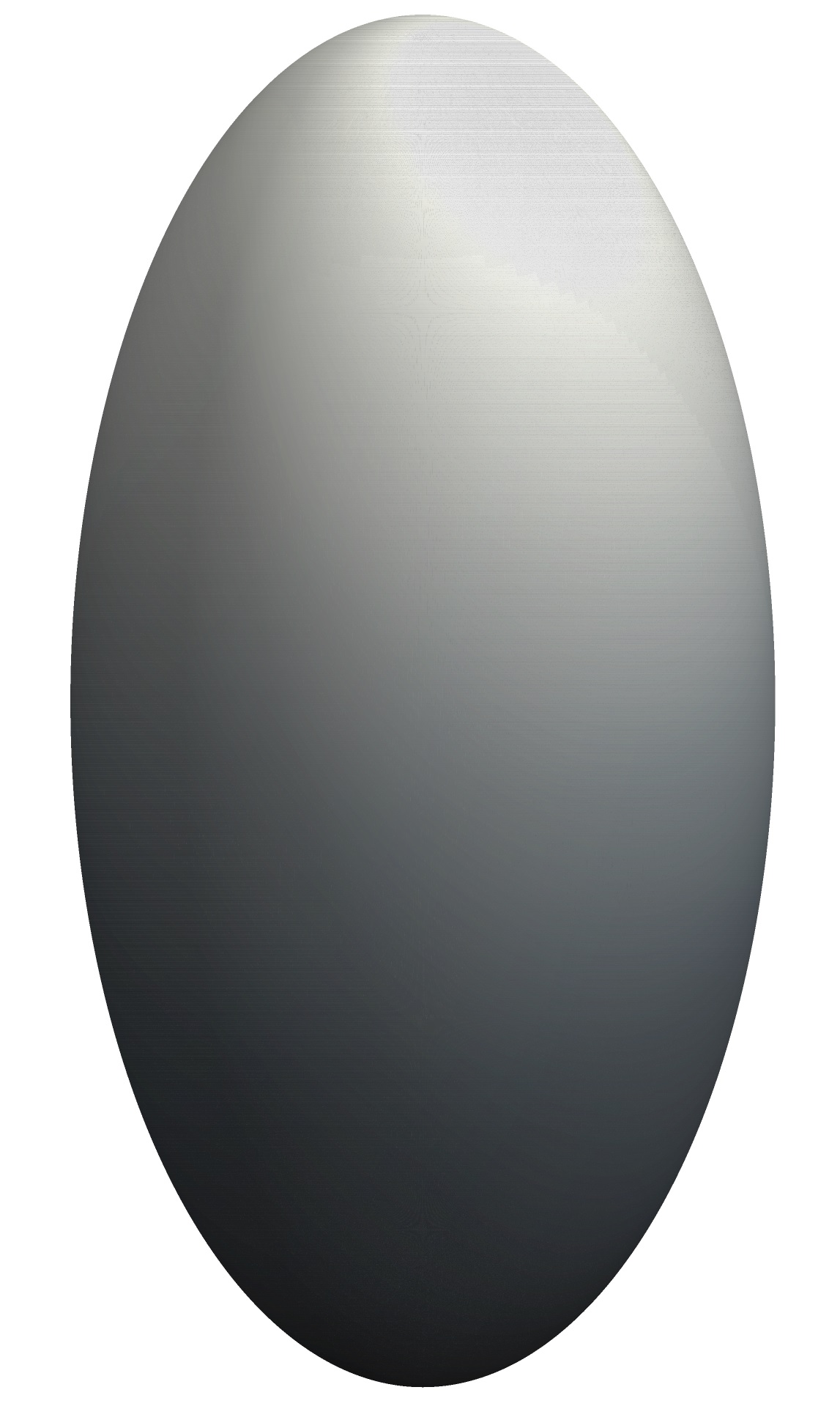} 
 \caption{The Schwarzschild electron temperature dependence on the polar angle $\theta$, Eq.~(\ref{SR_temp}), with $\kappa = 1$ and $s=0.9$.
 }
\label{Fig_SR_temp_dis}
\end{figure}

Now, let us turn to the spectral distribution of electromagnetic radiation from this accelerated charge.
While it is possible to derive the spectrum from first principles, it is easier to use the electron-mirror correspondence since the result for the mirror is known. This is the approach we take.

The single-sided beta Bogoliubov coefficient corresponding to the Schwarzschild remnant trajectory Eq.~\eqref{SR_traj} is given by  \cite{Good:2018zmx}
\begin{equation}
    \beta^{s}_{pq}=-\frac{s\sqrt{pq}}{2\pi\kappa(p+q)}\left(\frac{i\kappa}{p+q}\right)^{A}\Gamma(A),
\end{equation}
where $A=\frac{i}{2\kappa}[(1+s)p+(1-s)q]$, and $(p,q)$ are the frequencies in the mirror picture. Its modulus squared is
\begin{equation}
    \big|\beta^{s}_{pq}\big|^2=\frac{s^2 pq(p+q)^{-2}}{\pi\kappa(p+q+s(p-q))}\frac{1}{e^{\frac{\pi (p+q+s(p-q))}{\kappa}}-1}.
\end{equation}
Converting the mirror frequencies $(p,q)$ to the electron degrees of freedom $(\omega,\theta)$ (frequency and polar angle) and using the electron-mirror correspondence Eq.~\eqref{recipe_dIdOmega_from_mirror}, we obtain the spectral distribution
%
\begin{equation}
    \frac{dI}{d\Omega}=\frac{e^2s^2 \sin^2\theta}{16\pi^2\kappa (1+s\cos\theta)}\frac{\omega}{e^{\frac{\pi \omega(1+s\cos\theta)}{\kappa}}-1}\label{SR_dis}
\end{equation}
See Figure \ref{Fig_SR_spec_dis} for a spherical plot of the spectral distribution.  
It is straightforward to find the angular energy distribution is analytic,
\begin{equation}
    \frac{dE}{d\Omega}=\int \frac{dI}{d\Omega}d\omega=\frac{e^2s^2\kappa\sin^2\theta}{96\pi^2(1+s\cos\theta)^3}.\label{SR_energy_dis}
\end{equation}
See Figure \ref{Fig_SR_energy_dis} for a spherical plot of the energy distribution.  

The total energy can be found analytically,
\begin{equation}
    E=\int \frac{dE}{d\Omega}d\Omega=\frac{e^2\kappa}{24\pi}\left(\gamma^2-\frac{1}{s}\tanh^{-1}s\right).\label{SR_energy}
\end{equation}
As anticipated and required, this agrees with the result from the Larmor power Eq.~\eqref{SR_energy0}.
See Figure \ref{Fig_tot_energy} for a plot as a function of final speed $s$. 
In the limit $s\rightarrow 0$, the total energy vanishes since the trajectory, in this case, is static for all times; in the corresponding mirror, there will be no Doppler shift.
In the opposite limit $s\rightarrow 1$, the total energy diverges, as this corresponds to the usual eternal Schwarzschild black hole.

The spectral distribution  Eq.~\eqref{SR_dis} has an explicit Planck factor with the temperature
\be T = \frac{\kappa}{\pi(1+s\cos\theta)},\label{SR_temp}\ee
This temperature is angle-dependent; this is known to occur in three-dimensional moving mirrors, see e.g. \cite{Lin:2021bpe,PhysRevD.103.025014}.
The minimal and the maximal values are realized respectively at $\cos\theta \to 1$, and maximum, $\cos\theta \to -1$, 
\be T_\textrm{min} = \frac{\kappa}{\pi(1+s)} \to \frac{\kappa}{2\pi},\quad T_\textrm{max} = \frac{\kappa}{\pi(1-s)} \to \infty,\label{highspeedhightemp}\ee
The arrows here indicate the ultra-relativistic limit of $s\to 1$. 
This highlights the possibility of detecting photons distributed at a high temperature. See Figure \ref{Fig_SR_temp_dis} for an illustration of the temperature distribution, Eq.~(\ref{SR_temp}).

To recap, we have shown an electron accelerating along the interesting worldline Eq.~(\ref{SR_traj}), emitting a total finite energy Eq.~(\ref{SR_energy0}), has an associated Planck factor in the spectral distribution Eq.~(\ref{SR_dis}) and corresponding temperature, Eq.~(\ref{SR_temp}). The surprise is that the radiation can be arbitrarily hot, $T_{\textrm{max}} \to \infty$ as $s\to 1$, given by the second expression in Eq.~(\ref{highspeedhightemp}), observing the electron at $\theta = \pi$ in the ultra-relativistic limit.  


\section{CGHS}
\label{sec:CGHS_trajectory}

The Callan-Giddings-Harvey-Strominger (CGHS) system \cite{Callan:1992rs} is a two-dimensional theory of gravity that arises in the near-horizon approximation for a dilatonic black hole.
The corresponding metric can be cast in the form
\begin{equation}
    ds^2 = F(r) dt^2 - \frac{1}{F(r)} dr^2
\end{equation}
with
\begin{equation}
    F(r) = 1 - \frac{M}{\Lambda} e^{ - 2 \Lambda r } \,.
\end{equation}
A moving mirror dual to this black hole geometry was constructed and studied in \cite{Myrzakul:2021bgj}.

In this section, we consider an accelerated electron moving on the CGHS trajectory.
For clarity, we start with the black-hole dual case, where the electron approaches the speed of light (final speed $s=1$).  Then, we present new results on the remnant case, which has no horizon and a subliminal final speed of the electron, $s < 1$.

\subsection{CGHS black hole dual}

The 1+1d mirror trajectory dual to the CGHS black hole was derived in \cite{Myrzakul:2021bgj} (see Eq.~(14), (105), and (112) there and references therein).
In terms of the spatial coordinate $z$ and time $t$, it reads 
\begin{equation}
	z(t) = - \frac{1}{\kappa} \sinh[-1](\frac{ e^{\kappa t} }{2}).
\label{cghs_traj}
\end{equation}
This trajectory is determined by a single parameter $\kappa$. 
The motion starts asymptotically static, while at $t \to +\infty$, it approaches the speed of light asymptotically.

Now, let us consider an electron accelerated along the CGHS trajectory, Eq.~(\ref{cghs_traj}), and examine the radiation emitted. 
Again, the radiation spectral distribution can be derived from first principles; however, since the Bogolubov coefficients for this mirror are known, we will exploit the electron-mirror correspondence, Eq.~\eqref{recipe_dIdOmega_from_mirror},  as was done in the Schwarzschild case.

The Bogolubov coefficients for the CGHS moving mirror are given by (see Eq.~(19) of \cite{Myrzakul:2021bgj})
\begin{equation}
	\left|\beta^R_{pq}\right|^2
		= \frac{e^{\frac{2 \pi p}{\kappa}} \left( e^{\frac{2 \pi q}{\kappa} }-1 \right)}
			{2 \pi \kappa \left(p+q\right) \left(e^{\frac{2 \pi p}{\kappa}}-1\right)\left(e^{\frac{2 \pi (p+q) }{\kappa}}-1\right)} .
\end{equation}
Here, $p$ and $q$ are the frequencies of the incoming and outgoing modes, respectively ($\omega$ and $\omega'$ resp. in the notation of \cite{Myrzakul:2021bgj}).
Applying the prescription Eq.~\eqref{recipe_dIdOmega_from_mirror} we obtain the spectral distribution of the electron's radiation; it can be decomposed into two terms showing a distinct two-temperature spectrum:
\be
\frac{\diff{I}}{\diff{\Omega}} =
        \frac{  e^2 \omega  }{ 8 \kappa \pi^2 }\left[\frac{1}{e^{ \pi (1 + \cos\theta ) \omega / \kappa} - 1}
                - \frac{1}{e^{2 \pi \omega / \kappa} - 1}
            \right].\label{CGHS_two_plancks}
            \ee
The explicit Planck factor form of Eq.~(\ref{CGHS_two_plancks}) helps us disentangle the relevant thermal distribution directional dependence.  
%
%
For example, consider an observer positioned far from the origin at $\theta \sim \pi$ (but not exactly $\pi$); this is called the blueshift-forward limit since the electron is moving roughly towards the observer, cf. Table~\ref{tab:limits}.
In this case, the first term of Eq.~(\ref{CGHS_two_plancks}) dominates at high frequencies, so that
\be
\frac{\diff{I}}{\diff{\Omega}} \approx
        \frac{  e^2 \omega  }{ 8 \kappa \pi^2 }\frac{1}{e^{ \pi (1 + \cos\theta ) \omega / \kappa} - 1}. \label{cghsPlanck}\ee  
%
Around the forward direction, the spectral distribution is a Planck distribution with the temperature
\begin{equation}
	T_\theta = \frac{\kappa}{\pi  (1 + \cos\theta)}\label{CGHS_temperature} \,.
\end{equation}
The result Eq.~(\ref{cghsPlanck}) demonstrates the CGHS electron trajectory has a thermal signature of Planck-distributed photons as analogously expected from the thermal CGHS black hole.  

The presence of an event horizon is apparent from the fact that the spectrum, using Eq.~(\ref{CGHS_two_plancks}),
\begin{equation}
	I(\omega) = \int \diff\Omega \frac{\diff{I}}{\diff{\Omega}},
\end{equation}
is divergent. 
Consequently, the total radiated energy (and photon count) is infinite.
This will be remedied in the next subsection by introducing a final subliminal speed $s$.

\subsection{CGHS remnant dual}

While the CGHS electron trajectory has thermal emission as evidenced by the Planck factor, Eq.~(\ref{cghsPlanck}), it also has a horizon. The spectrum $I(\omega)$ for the electron does not exist (only the spectral distribution $\diff{I}/\diff{\Omega}$, Eq.~(\ref{CGHS_two_plancks}), makes sense).

We will now deform the trajectory, Eq~(\ref{cghs_traj}), to remove the horizon by introducing a final speed $s < 1$.
This modification can be viewed as a regularization. 
We will investigate the spectrum and check the correct distribution limits in the $s \to 1$ regime.

\subsubsection{Trajectory and global properties}

\begin{figure}[h]
\includegraphics[width=\columnwidth]{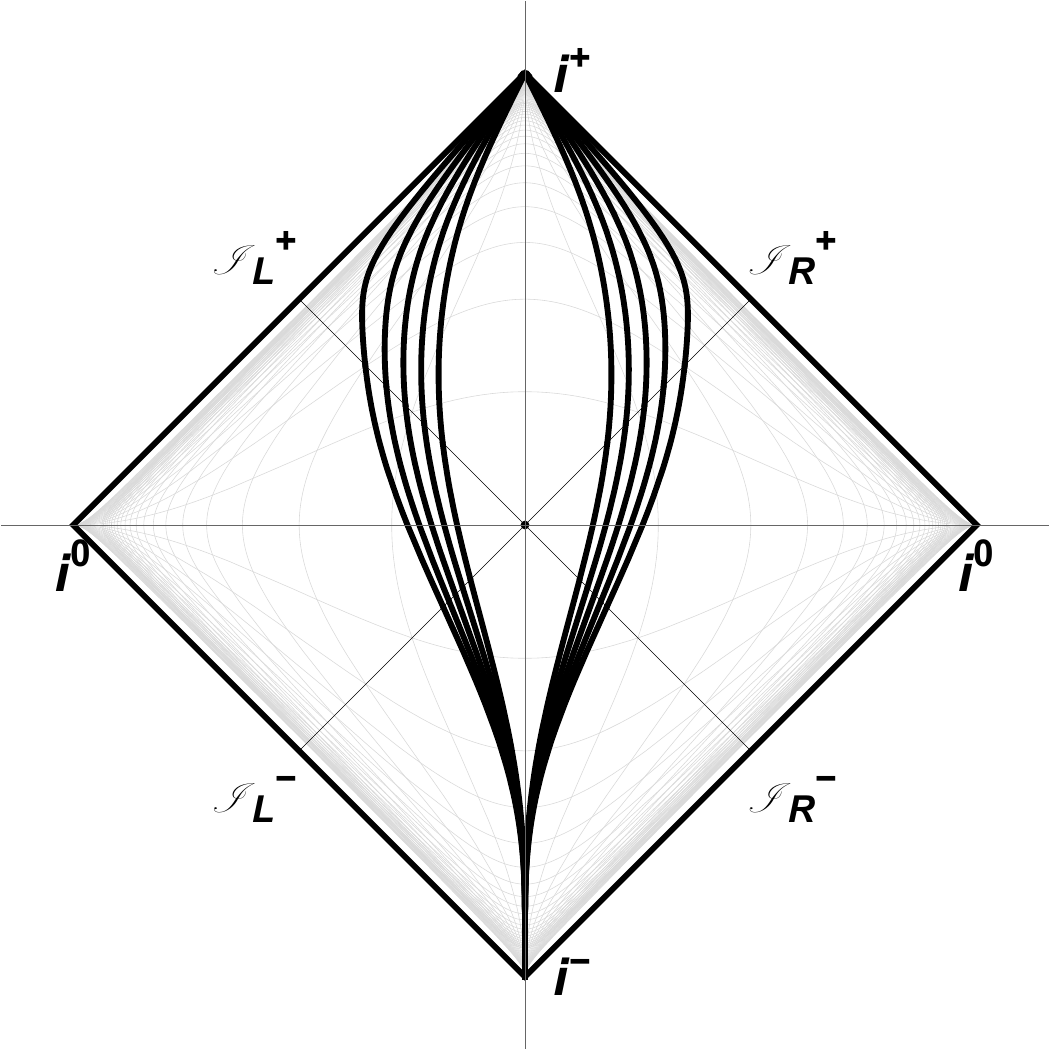} 
 \caption{A Penrose conformal diagram of the CGHS remnant trajectory, Eq.~(\ref{cghs_traj_deformed}) traveling both to the left or right.  Notice the worldlines are time-like, especially in the asymptotic future.  Here $\kappa =1$ and the final speeds are $s=0.5, 0.6, 0.7, 0.8, 0.9$ from the inside-out.  Here, only one spatial direction is plotted in the horizontal direction, while time is in the vertical direction. Similar plots can be constructed for the Schwarzschild remnant trajectory and other asymptotic constant velocity worldlines.  
 }
\label{Fig_Penrose_CGHS_R}
\end{figure}

This deformation can be introduced straightforwardly for a trajectory in the form $z = z(t)$ Eq.~\eqref{cghs_traj}: it is sufficient to rescale the space coordinate by a factor of $s$ and write
\begin{equation}
	z(t) = - \frac{s}{\kappa} \sinh[-1](\frac{ e^{\kappa t} }{2}).
\label{cghs_traj_deformed}
\end{equation}
This is our proposed CGHS remnant trajectory; it can easily be checked that the speed approaches $s$ at late times.
See Figure \ref{Fig_Penrose_CGHS_R} for a conformal diagram depicting the lack of a horizon and an asymptotic constant velocity end-state. 

The peel acceleration for this trajectory is readily found from $\bar{\kappa} = 2\alpha e^{\eta}$ where $\alpha$ and $\eta$ are the proper acceleration and rapidity, respectively,
\be \bar{\kappa} = 8 \kappa  s e^{\kappa  t}\frac{ s e^{\kappa  t}-\sqrt{e^{2 \kappa  t}+4}}{\left(\left(s^2-1\right) e^{2 \kappa  t}-4\right)^2}.\label{CGHS_peel}\ee
In the regime of fast speeds and late times, the peel is constant, which is indicative of thermality \cite{Ievlev:2023inj},
\be \lim_{t\to\infty} \lim_{s\to 1} |\bar{\kappa}| = \kappa.\ee
See Figure \ref{Fig_Peel} for a plot of the CGHS peel acceleration as a function of coordinate time $t$, and the associated plateau, which signals a thermal emission.  This plateau suggests that the CGHS remnant spectral distribution will have a thermal signature despite the trajectory having no horizon and finite energy emission.  

To compute this total radiated energy, consider the radiated power in terms of the proper acceleration $\alpha$, 
\be P = \frac{e^2\alpha^2}{6\pi} = \frac{8 e^2 \kappa ^2 s^2 e^{2 \kappa  t}}{3 \pi  \left(\left(1-s^2\right) e^{2 \kappa  t}+4\right)^3}.\ee
The total energy is computed as an integral of the power over time,
\be 
E = \int_{-\infty}^{\infty} P(t) \diff{t} = \frac{e^2\kappa}{24\pi} \left[\gamma_s^2 -1\right].\label{CGHS_energy}
\ee
The main takeaway is the total radiated energy is finite, unlike the radiation emitted from the $s=1$ CGHS trajectory Eq.~(\ref{cghs_traj}). The total energy, Eq.~(\ref{CGHS_energy}), is proportional to the final celerity (proper velocity) squared, $w_s^2 = \gamma_s^2 s^2 = \gamma_s^2-1$. 

The Feynman power can further confirm Eq.~(\ref{CGHS_energy}).  The self-force, $F = e^2\alpha'(\tau)/6\pi$, multiplied by velocity, $\diff{r}/\diff{t}$, gives the Feynman power \cite{Feynman:1996kb}, 
\be
F\cdot v = -\frac{4 e^2 \kappa ^2 s^2 e^{2 \kappa  t} \left(\left(1-s^2\right) e^{2 \kappa  t}-2\right)}{3 \pi  \left(\left(1-s^2\right) e^{2 \kappa  t}+4\right)^3},\ee
whose radiation reaction is consistent with energy conservation of Eq.~(\ref{CGHS_energy}),
\be E = -\int_{-\infty}^{\infty} F \cdot v \diff{t}= \frac{e^2\kappa}{24\pi} \left[\gamma_s^2 -1\right].\label{CGHS_energy1}\ee

Using spectral analysis in the moving mirror model, we can also confirm the total energy emitted, Eq.~(\ref{CGHS_energy}) and Eq.~(\ref{CGHS_energy1}).  
The spectrum as found via both sides of the moving mirror is (keep in mind there is no electric charge)
\be
|\beta_{pq}|^2 = \frac{8 s^2 p q e^{\frac{\pi c}{\kappa }} Q}{\pi  a b c \kappa  \left(e^{\frac{\pi  a}{\kappa }}-1\right) \left(e^{\frac{\pi  b}{\kappa }}-1\right) \left(e^{\frac{\pi  c}{\kappa }}-1\right)}.\label{CGHS_betas}
\ee
Here $Q = \cosh \left(\frac{\pi  a}{\kappa }\right)+\cosh \left(\frac{\pi  b}{\kappa }\right)-2$.  We use $a= p(1+s) + q(1-s)$, $b=p(1-s)+q(1+s)$, and $a + b = c$. Here $a-b = d$. Note that $c = 2(p + q)$.  Again, the final drifting speed is labeled $s$, and always $1>s>0$. A numerical integration,
\be E = \int_0^\infty\int_0^\infty p|\beta_{pq}|^2 \diff{p}\diff{q} = \frac{\kappa}{24\pi}[\gamma_s^2 - 1],\ee
demonstrates Eq.~(\ref{CGHS_betas}) is analog-consistent with the total energy emitted as given by Eq.~(\ref{CGHS_energy}).
This consistency illustrates a powerful analog between the mirror and electron systems.

\subsubsection{Electromagnetic spectrum \& temperature}

\begin{figure}[h]
\includegraphics[width=0.6\linewidth]{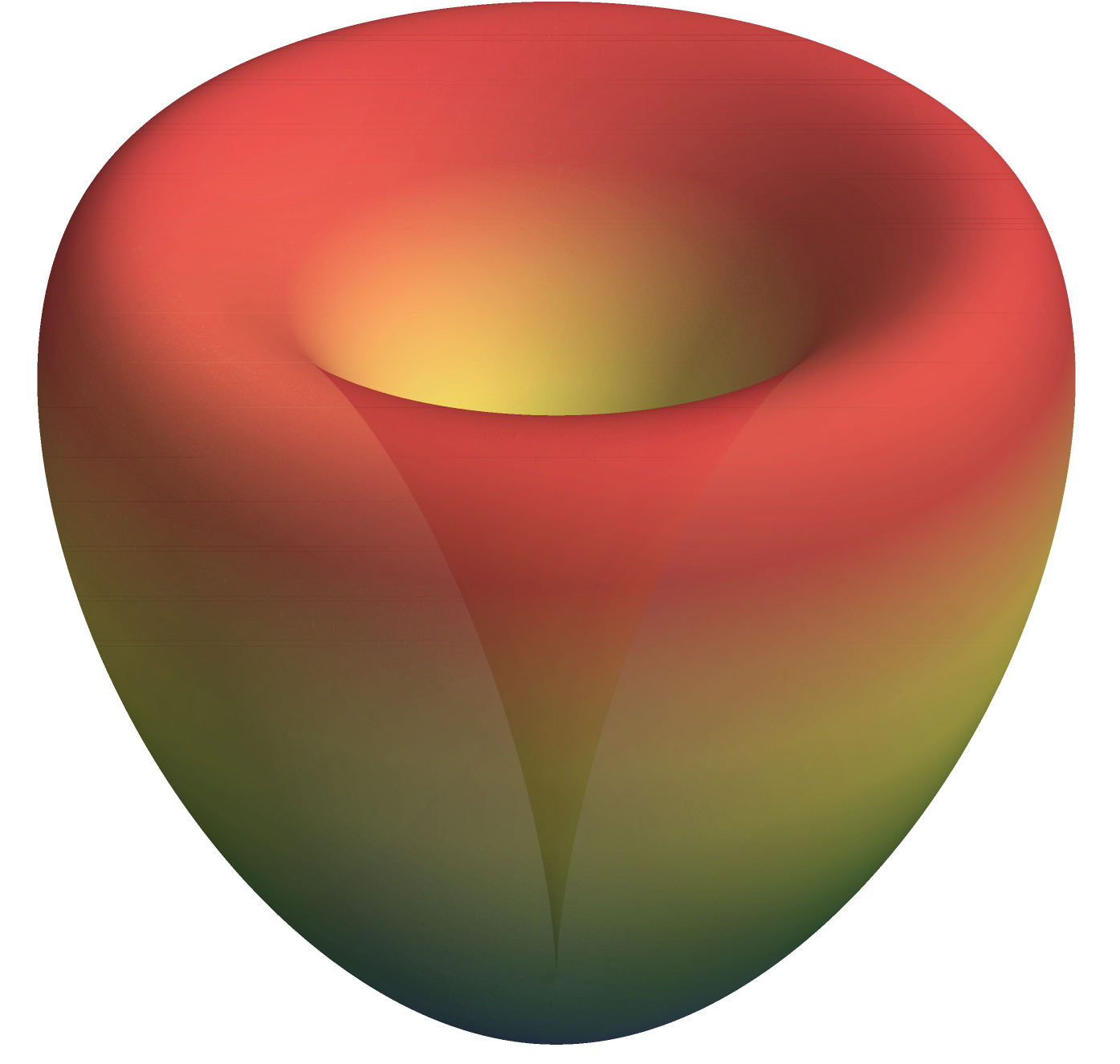} 
 \caption{A spherical plot of the spectral distribution, Eq.~(\ref{cghs_deformed_dIdOmega}). Here $\omega = \kappa =1$ and $s = 0.9$.  
 }
\label{specdis}
\end{figure}

\begin{figure}[h]
\includegraphics[width=0.6\linewidth]{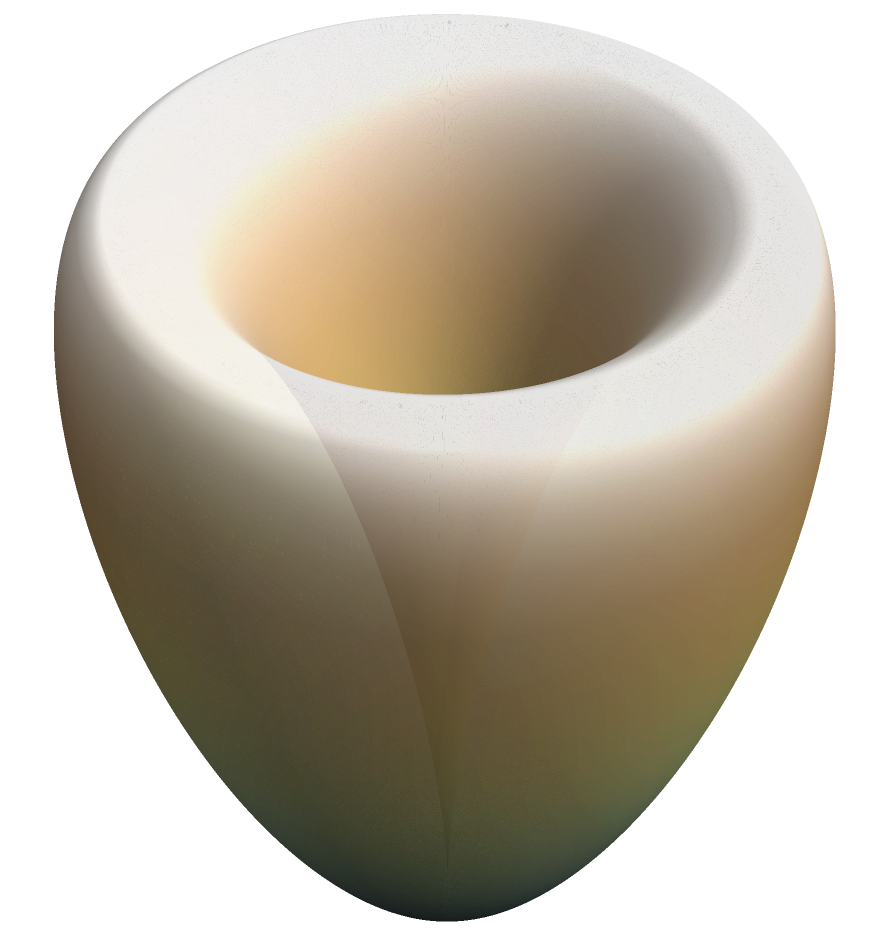} 
 \caption{A spherical plot of the energy distribution, Eq.~(\ref{E_omega}). Here $\kappa =1$ and $s = 0.9$.  
 }
\label{energy3D}
\end{figure}

\begin{figure}[h]
\includegraphics[width=\columnwidth]{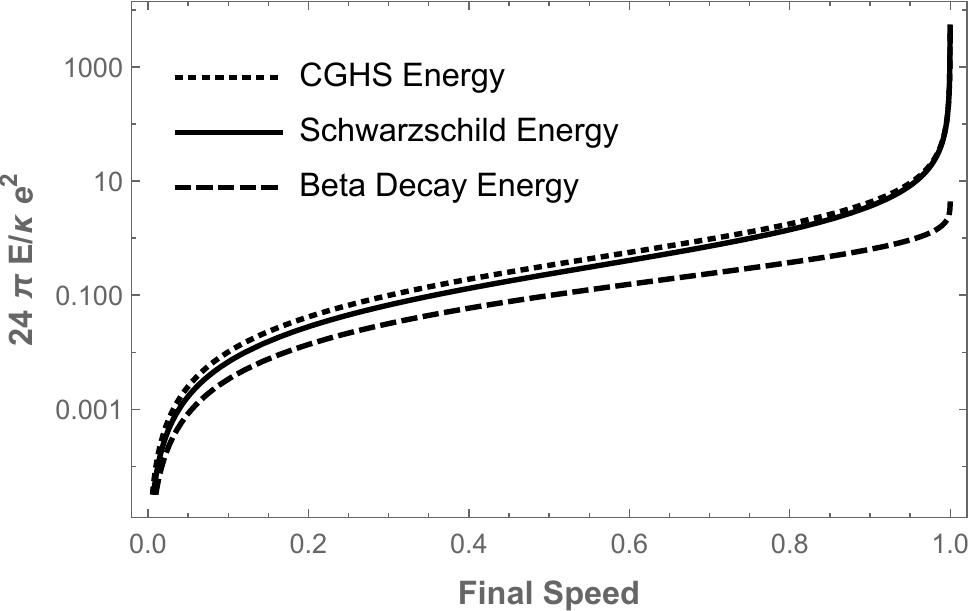} 
 \caption{A log plot of the total energy emitted by three electron trajectories, including the Schwarzschild, Eq~(\ref{SR_energy}), CGHS, Eq.~(\ref{CGHS_energy}), and the beta decay trajectory, e.g. Eq. 7 of \cite{Ievlev:2023inj}; see also Eq. 3 of \cite{Good:2022eub}.  Here, all three trajectories are normalized by $24\pi/\kappa e^2$, and the final coasting speed ranges from $0<s<0.9999$ for illustration. Key takeaways are the finite energy emission, monotonic increase, and minimal infrared lower bound of beta decay.
 }
\label{Fig_tot_energy}
\end{figure}

The spectral distribution for an electron moving on the trajectory Eq.~\eqref{cghs_traj_deformed} can be found by applying the aforementioned recipe, and the result is only a small modification of the dual-temperature spectral distribution of Eq.~(\ref{CGHS_two_plancks}): 
\begin{equation}
\begin{aligned}
    \frac{\diff{I}}{\diff{\Omega}} &=
        \frac{ s^2 e^2 \omega  }{ 8 \kappa \pi^2 }  \,
            \frac{ \sin^2\theta }{ 1 - s^2 \cos^2\theta } \\
            &\times
            \left[ 
                \frac{1}{e^{ \pi (1 + s \cos\theta ) \omega / \kappa} - 1}
                - \frac{1}{e^{2 \pi \omega / \kappa} - 1}
            \right].
\end{aligned}
\label{cghs_deformed_dIdOmega}
\end{equation}
In Fig.~\ref{specdis}, we show a spherical plot of the spectral distribution to highlight the 3+1 dimensionality of the result, as opposed to the 1+1 dimensional moving mirror model. 

Integrating Eq.~(\ref{cghs_deformed_dIdOmega}) over the frequency $\omega$ gives the angular energy distribution,
\begin{equation}
    \dv{E}{\Omega}
        = \int \frac{\diff{I}}{\diff{\Omega}}\diff{\omega} 
        = \frac{e^2 \kappa s^2 (3 + s \cos\theta) \sin^2\theta }{ 192 \pi^2 (1 + s \cos\theta)^3 },
\label{E_omega}
\end{equation}
Figure \ref{energy3D} illustrates this CGHS energy angular distribution.
The total energy is then computed by integrating Eq.~\eqref{E_omega} over the solid angle $\diff{\Omega} = \sin\theta \diff{\theta}\diff{\phi}$,
\be E = \int \dv{E}{\Omega} \diff{\Omega} = \frac{e^2\kappa s^2 }{24\pi (1-s^2) },\ee
This is the total energy formula Eq.~\eqref{CGHS_energy} and confirms the correctness of the spectral analysis. It also confirms the mirror-electron recipe. 
See Figure \ref{Fig_tot_energy} for a plot of the total energy as a function of the final speed $s$.

In the blueshift-forward limit $\theta \to \pi$ at high final speeds $s \to 1$ (but $s \neq 1$ exactly), the spectral distribution Eq.~\eqref{cghs_deformed_dIdOmega} is again dominated by the first term, which yields pure Planck spectrum with the temperature
\begin{equation}
	T = \frac{\kappa}{\pi(1+s \cos\theta)} \,.
\end{equation}
There is a significant physical difference when $s < 1$.  Now the temperature is finite even at exactly $\theta = \pi$, in contrast to the temperature Eq.~(\ref{CGHS_temperature}) of the horizon trajectory Eq.~(\ref{cghs_traj}).

To recap, the CGHS electron trajectory Eq.~(\ref{cghs_traj}) emits thermal radiation as evidenced by the Planck factor, Eq.~(\ref{cghsPlanck}), but also has a horizon that spoils the energy emission (infinite).  It also produces a singularity in the temperature when measured at $\theta = \pi$. The deformed trajectory, Eq~(\ref{cghs_traj_deformed}), removes the horizon by introducing a final speed $s < 1$ and fixes the energy emission (finite).  Moreover, it regularizes the temperature and results in a finite measurement at $\theta = \pi$.  The electron-mirror correspondence consistently provides the correct result for the finite amount of radiated energy (as well as spectral and time-dependent quantities).  The most physically unexpected result is the range of temperatures that the remnant system may exhibit, which are both speed and $\theta$-angle dependent.

\section{BTZ}
\label{sec:BTZ}

In this section, we consider a mirror that corresponds to a non-rotating BTZ black hole in 2+1 dimensional AdS spacetime \cite{Banados:1992wn}.
The spacetime metric, in this case, reads
\begin{equation}
\begin{aligned}
    ds^2 &= F(r) dt^2 - \frac{1}{F(r)} dr^2 - r^2 d\phi^2 \,, \\
    &F(r) = - M + \frac{r^2}{\ell^2} \,.
\end{aligned}
\label{in_out_metric_null_btz}
\end{equation}
Here, $\ell$ is a parameter of the dimension of length (related to the size of the AdS), while $M$ is the black hole mass (recall that in 2+1d, mass is dimensionless).

\begin{figure}[h]
    \centering
    \includegraphics[width=0.7\columnwidth]{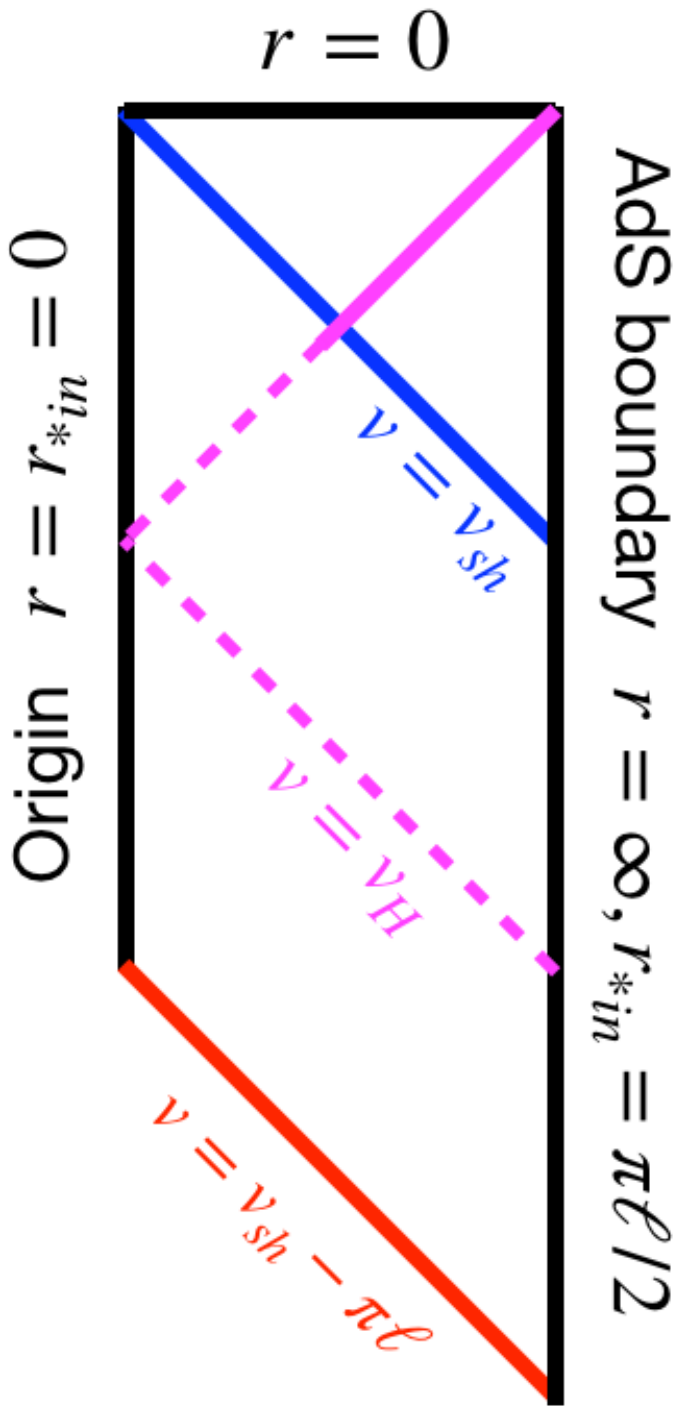}
    \caption{
    	Penrose diagram of the formation of a non-rotating BTZ black hole (region above $v=v_{\text{sh}}$) from the gravitational collapse of a null shell $(v=v_{\text{sh}})$ in AdS space (region below $v=v_{\text{sh}}$). Colored in magenta is the non-rotating BTZ black hole's event horizon. In this scenario, only ingoing waves emitted at $v\in(v_{\text{sh}}-\pi\ell,v_{\mathcal{H}})$ can be received by an observer at the timelike AdS boundary where $v>v_{\text{sh}}$.}
    \label{fig:btz_penrose}
\end{figure}

\subsection{Mirror dual to BTZ black hole}

The mirror corresponding to the geometry Eq.~\eqref{in_out_metric_null_btz} was worked out by one of the authors in \cite{Ievlev:2023ejs}.
For completeness, we briefly review some basic important facts, elaborate on essential points, and further extend the analysis.
In light-cone coordinates, the mirror trajectory is given by
\begin{equation}
	f_m(v) = v_\text{sh} + \frac{2\ell}{\sqrt{M}} \tanh[-1]( \frac{\sqrt{M}}{ \tan( \frac{v_\text{sh} - v}{2\ell} ) } )
\label{BTZ_f}
\end{equation}
The parameter $v_\text{sh}$ is the $v$-coordinate of a collapsing shell forming the black hole; one can always choose $v_\text{sh} = 0$ for simplicity.

If taken as-is, this trajectory is periodic in $v$ and, therefore, has infinite branches.
This is a manifestation of the fact that AdS spacetime admits closed timelike curves.  To rectify this, one must pass to the covering space, which in the mirror's language means that we should stick to any particular branch of the trajectory Eq.~\eqref{BTZ_f}.

Furthermore, the timelike boundary of AdS implies that in a null shell collapse model, a light ray emitted too early never actually encounters the collapsing shell.
Indeed, from the AdS tortoise coordinate
\begin{equation}
	r_{*\text{in}}(r) = \int\limits_0^r \frac{ \diff \rho }{ F_\text{in}(\rho) } = \ell \tan[-1](\frac{r}{\ell})
\label{BTZ_tortoise_inner}
\end{equation}
we see that a light ray emitted from the past null infinity reaches the origin in finite coordinate time $\Delta t_\text{in} = \pi\ell/2$,  gets \textquote{reflected}, and then reaches the future null infinity in another lapse $\Delta t_\text{in} = \pi\ell/2$, see Fig.~\ref{fig:btz_penrose}. 
Thus, it takes $\pi\ell$ time for a light ray to cross the whole of AdS.

Implications of this can be seen from Fig.~\ref{fig:btz_penrose}.
Take a collapsing null shell propagating at $v = v_\text{sh}$.
Light rays emitted from the past null infinity $\mathcal{I}^R_-$ too early, i.e. at $v < v_\text{sh} - \pi\ell$, never actually encounter the collapsing null shell.

All this means is that on the mirror side, one should not consider the values $v < v_\text{sh} - \pi\ell$.
The future horizon of the trajectory is defined by $f_m(v_\mathcal{H}) = + \infty$, which gives
\begin{equation}
	v_\mathcal{H} = v_\text{sh} - 2 \ell \tan[-1]( \sqrt{M} ) \quad
	\text{for } M>0
\label{btz_horizon_v}
\end{equation}
Thus, we finally arrive at the formula for the BTZ radiation spectrum,
\begin{equation}
	\beta^R_{pq}
		=\frac{1}{2 \pi} \sqrt{\frac{q}{p}} 
		\int\limits_{v_\text{sh} - \pi\ell}^{ v_\mathcal{H}  } d v \,
			e^{-i q v -i p f_m(v)}
\label{btz_betas}
\end{equation}
This prescription removes the fictitious past horizon.

Finally, we note that Eq.~\eqref{btz_betas} only gives the Hawking radiation from the black hole. 
It does not take into account possible radiation from AdS boundary.
In particular, if we remove the black hole completely and consider pure AdS$_3$, the beta Bogolubov coefficients will be trivial.
Indeed, AdS$_3$ can be viewed as the BTZ black hole with mass $M=-1$ \cite{Banados:1992wn}. Substituting $M=-1$ into the trajectory Eq.~\eqref{BTZ_f}, we obtain
\begin{equation}
	f_m(v) = - 2\ell \tan[-1]( \cot( \frac{v}{2\ell} )  )
		= v - \pi \ell
\label{BTZ_f_M=-1}
\end{equation}
Thus, we end up with a static mirror. Such a mirror does not produce radiation.

Now consider the proposed BTZ trajectory of Eq.~\eqref{BTZ_f}.
We want to compute the integral in Eq.~\eqref{btz_betas} to find the spectrum (as has been done for the Schwarzschild and CGHS systems); however, it is difficult to evaluate Eq.~\eqref{btz_betas} analytically.
Instead, we appeal to the near-horizon approximation, as is often done on the gravity side; see \cite{Hyun:1994na} for the case of the BTZ black hole.

The future horizon of the trajectory Eq.~\eqref{BTZ_f} is located at Eq.~\eqref{btz_horizon_v}.
In the near-horizon (or late-time) approximation, the trajectory Eq.~\eqref{BTZ_f} becomes
\begin{equation}
	f_m(v) \approx - \frac{\ell}{ \sqrt{M} } \ln(- (v - v_\mathcal{H}) )
\label{BTZ_f_ellhalf_nearhor}
\end{equation}
This form is not surprising at all.
In the near-horizon approximation, every mirror that corresponds to a black hole looks like Carlitz-Willey \cite{carlitz1987reflections}; one only has to identify the parameters correctly.
The beta integral can now be done with ease and leads to
\begin{equation}
	|\beta^R_{pq}|^2 \approx \frac{1}{2 \pi \kappa q} \frac{1}{ e^{2 \pi p / \kappa} - 1 } \,, \quad
	\kappa = \frac{\sqrt{M} }{ \ell }
 \label{BTZ-beta-identified}
\end{equation}
The Planck factor is explicit, and the  temperature is
\begin{equation}
	T_\text{BTZ} = \frac{\sqrt{M} }{ 2 \pi \ell }
\label{BTZ-temperature-mirror}
\end{equation}
These results agree with the calculation from QFT in curved spacetime, see e.g. \cite{Banados:1992wn,Hyun:1994na}.

\subsection{Mirror dual to BTZ remnant}

In the spirit of the other two cases considered in this paper, we would like to treat the case of a BTZ remnant.
We do this in two steps.

First, by taking a time derivative of a trajectory in the form $u = f_m(v)$, we can derive the velocity
\begin{equation}
    \dot{Z}(t) = \frac{1 + f_m'(v)}{1 - f_m'(v)} \,,
\end{equation}
where we have made use of $v=t+Z(t)$ and $f_{m}(v)=t-Z(t)$
Here, primes denote a derivative with respect to $v$, while the dot is the time derivative.
At late times (near the horizon) $f_m'(v)$ diverges, and the trajectory approaches the speed of light $|\dot{Z}| \to 1$.
We aim to deform the trajectory so that the final speed would be subliminal, $|\dot{Z}| \to s < 1$.

\begin{figure}[h]
\includegraphics[width=\columnwidth]{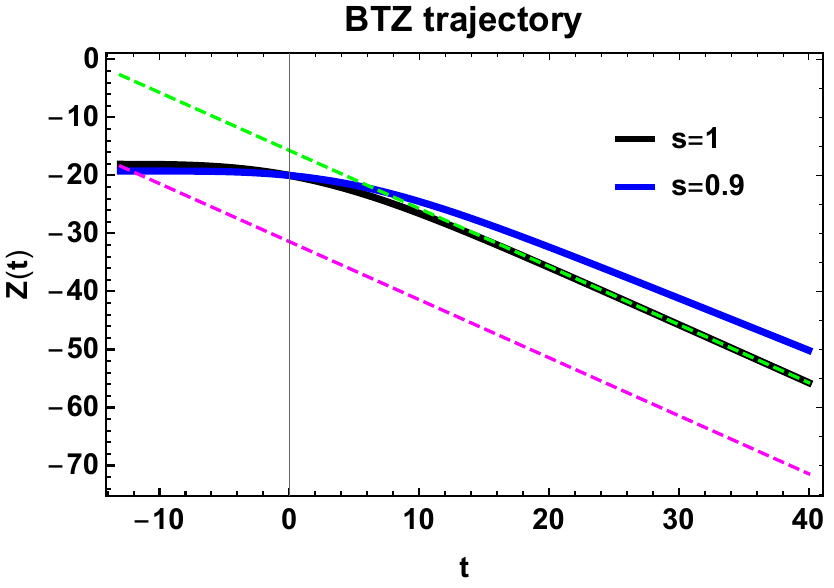} 
 \caption{BTZ trajectories. $M=1,\ell=10,v_{\text{sh}}=0$ and initial condition: $Z(0)=-20$. The black curve is asymptotically null (corresponding to a BTZ black hole), while the blue curve has a final speed $s=0.9$ (corresponding to a BTZ remnant). The green dashed line is $Z(t)=-t+v_{\mathcal{H}}$, and the magenta dashed line is $Z(t)=-t+v_{sh}-\pi\ell$. For $s=1$ (horizon present), the trajectory (black) is bounded since $v_{sh}-\pi\ell<v<v_{\mathcal{H}}$.
 }
\label{Fig_BTZ-trajectory}
\end{figure}

In our case, using the explicit trajectory Eq.~\eqref{BTZ_f}, we can obtain the velocity as
%
\begin{equation}
    \dot{Z}(t) =\frac{(1+M) + (1+M)\cos\big(\frac{t+Z(t)-v_{\text{sh}}}{\ell}\big)}{-3+M+(1+M)\cos\big(\frac{t+Z(t)-v_{\text{sh}}}{\ell}\big)} \,.
\label{btz_z_of_t}
\end{equation}
%
Eq.~\eqref{btz_z_of_t} corresponds to a BTZ black hole with an event horizon, and at late times, it approaches the speed of light.
To get a new trajectory that has asymptotic speed $s$, we deform Eq.~\eqref{btz_z_of_t} by substituting $Z(t)\rightarrow Z(t)/s$.
For trajectories with a final speed $s$, the velocity can be obtained by substituting $Z(t)\rightarrow Z(t)/s$ in the above formula. Then,
%
\begin{equation}
\begin{aligned}
    \dot{Z}(t) &= \frac{s(1+M) ( 1 + \cos \zeta ) }{-3+M+(1+M)\cos \zeta } \,, \\
    \zeta &\equiv \frac{t+Z(t)/s-v_{\text{sh}}}{\ell} \,.
\end{aligned}
\label{BTZ-remnant-velocity}
\end{equation}
For a comparison with the undeformed trajectory, see Fig.~\ref{Fig_BTZ-trajectory}.  We can now examine the radiation from an accelerated electron analogous to a BTZ black hole and BTZ remnant.

\subsection{Electron dual to BTZ}

Let us discuss the electron dual to the BTZ black hole and the remnant.
The corresponding electron trajectories for these systems are the same as the mirror trajectories Eq.~\eqref{BTZ_f} and Eq.~\eqref{BTZ-remnant-velocity}.

\subsubsection{Electron dual to BTZ black hole}

In the case of the BTZ black hole, the late-time (near-horizon) approximation on the gravity/mirror side corresponds to the late-time (ultrarelativistic) approximation on the electron side.
Using the electron-mirror recipe Eq.~\eqref{recipe_dIdOmega_from_mirror}, we transform the Bogolubov coefficients from Eq.~\eqref{BTZ-beta-identified} to obtain the spectral distribution of the accelerated electron's radiation,
\begin{equation}
    \dv{I(\omega)}{\Omega}
        \approx \frac{\omega}{ 4 \pi^2 \kappa (1 - \cos\theta) } \frac{1}{e^{\frac{\pi \omega (1 + \cos\theta)}{ \kappa}} - 1} \,.
\end{equation}
Immediately, we find the Planck spectrum with temperature,
\begin{equation}
    T_\theta = \frac{ \kappa }{ \pi (1 + \cos\theta) } \,.
\label{BTZ-temperature-electron}
\end{equation}

Comparing the electron temperature Eq.~\eqref{BTZ-temperature-electron} with the mirror/BH temperature Eq.~\eqref{BTZ-temperature-mirror} we have rigorously confirmed the same thermal phenomenon: the temperature is angle-dependent, and for an observer in front of this accelerated electron (i.e. at $\theta = \pi$, since the electron moves in the direction $z \to - \infty$), the temperature becomes unphysically singular.

\subsubsection{Electron dual to BTZ remnant}

\begin{figure}[h]
\includegraphics[width=\columnwidth]{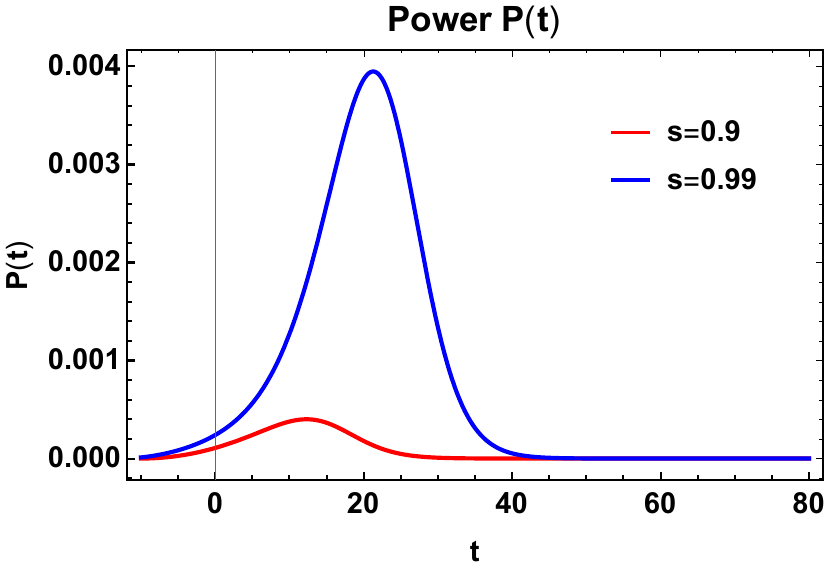} 
 \caption{Larmor power for BTZ trajectories. $M=1,\ell=10,v_{\text{sh}}=0,e=1,$ and initial condition: $Z(0)=-20$. The total energies are finite for $s<1$.
 }
\label{Fig_BTZ-power}
\end{figure}

To remedy this problem, let us consider the BTZ remnant.
Having an explicit expression for the velocity Eq.~\eqref{BTZ-remnant-velocity}, it is straightforward to write down the Larmor power $P =e^2\alpha^2/6\pi$.
The expression is quite bulky, so we do not present it here; for a plot, see Fig.~\ref{Fig_BTZ-power}.

Based on the experience gained from the Schwarzschild and CGHS cases, the total energy is expected to have a general form
\begin{equation}
    E=\int_{-\infty}^{\infty} P(t)dt=\frac{e^2\kappa}{24\pi} \left( \frac{1}{1-s^2}-g(s) \right)  \,,
\end{equation}
where $g(s)$ is a function of the final speed $s$ that depends on the particular shape of the trajectory, i.e., how the remnant forms.
The first term represents a pole contribution in the light-speed limit $s \to \pm 1$.
This term is universal for BH/remnants, since at $s \to \pm 1$, a remnant trajectory turns into an un-regularized black hole trajectory, which, by the no-hair theorem, does not depend on the detail of the formation.

Other types of singularities in the total energy also appear in the literature.
For example, the trajectory studied in \cite{Good:2022eub,Ievlev:2023inj} has a branch point singularity in the total energy, $E \sim \ln(1 \pm s)$.
While it is beyond the scope of this work, it would nevertheless be interesting to understand what geometry corresponds to these trajectories.

The temperature of the BTZ remnant system is given by the deformed version of Eq.~\eqref{BTZ-temperature-electron},
\begin{equation}
    T_\theta = \frac{ \kappa }{ \pi (1 + s\cos\theta) } \,.
\end{equation}
In the blueshift-forward limit $\theta \sim \pi$ the temperature is now finite,
\begin{equation}
    T_\text{b.f.} = \frac{\kappa}{\pi (1 - s) } \,.
\label{BTZ-rem-temperature-electron}
\end{equation}
For a very fast subliminal speed $s\sim 1$ final velocity, the radiation emitted by the electron can be arbitrarily high but finite.  This starkly contrasts with the corresponding moving mirror temperature (BTZ black hole temperature).

\section{Conclusions} 
\label{sec:conclusions}

We have explored the triality between accelerated electrons, moving mirrors, and black holes. The investigation focused on this correspondence for the case of black hole remnants and the implication for the analog behavior of the radiation emitted by their corresponding analogs as accelerated electrons.
The three examples covered here --- Schwarzschild, CGHS, and BTZ --- share some common properties; let us summarize them here.

First, the total radiated energy $E$ is finite for the remnant regularization cases.  This contrasts with the un-regularized infinite evaporation energy of the black hole cases (e.g. electron analog sets $s=1$).
On the electron side, the total energy depends on the asymptotic final speed $s$ and has an expected pole in the lightspeed limit,
\begin{equation}
    E = \frac{e^2\kappa}{24\pi} \frac{1}{ 1 - s^2 } + \ldots \,,
\end{equation}
where the dots stand for finite terms.

Second, we have found that classical electromagnetic radiation from the accelerated electron has a Planck-distributed spectral form.  The associated temperature can be much higher than the corresponding black hole or mirror.
Generally speaking, the electron temperature, observed at spatial infinity (in the blueshift-forward limit, when the electron flies towards the observer) is given by
\begin{equation}
    T_\text{el} = \frac{ 2T_\text{BH/mir} }{ 1 - s } \,,
\end{equation}
where $s$ is the asymptotic final speed of the electron, and $T_\text{BH/mir}$ is the temperature of the corresponding black hole or asymptotically null mirror (the latter two are the same).
These high-speed electrons have high temperatures; this can be contrasted with the thermal trajectory associated with radiative beta decay studied in \cite{Good:2022eub,Ievlev:2023inj}, where the temperature depends only on the acceleration (or bandwidth of the allowed frequencies).  
This implies that systems with totally different temperature scales may be actually related via the electron-mirror duality. 

Our calculation shows the finite total radiated energy in the remnant cases.  One can ask a question: what is the final mass of this remnant?
Since we do not incorporate the gravitational potential and greybody factors in this model, currently, we can put only the upper bound as the black hole mass $M$ minus the total Hawking radiated energy.
The latter can be easily computed on the electron side with the help of the Larmor formula.
Accounting for the parameters of the correspondence (see Table~\ref{role_of_parameters}), we have in SI units
\begin{equation}
    m_\text{rem} c^2 \lesssim M c^2 - E_\text{Hawk} = M c^2 - \frac{ \hbar }{ \mu_0 c e^2 } E_\text{electron}
\label{remnant_mass_estimate}
\end{equation}
Since we do not include complications like backscattering, the estimate Eq.~\eqref{remnant_mass_estimate} is valid only while the radiated energy is much smaller than the initial mass, $E_\text{Hawk} \ll M c^2$.
In summary, exact classical computation reveals thermal radiation emitted by a moving point charge in several relevant black hole geometries related to remnants. The systems possess a speed-dependent temperature that characterizes the emission from the accelerated electron.  Exploiting the classical-quantum correspondence between the electron and moving mirror allows an efficient approach to analytically determining the spectra. 

Our analysis relies on time dependence, and these regularized (i.e., subliminal velocity) dynamics ultimately determine the temperature of the remnant analog and its finite radiation energy. The approach has similarities to previous arguments  \cite{Adler:2001vs}, based on dynamics, for the existence of a black hole remnant, which also radiates a finite energy.  An interesting future study could investigate the finite energy, Eq.~(\ref{SR_energy0}), and its similarity to the finite energy output\footnote{Eq. (7b) of \cite{Adler:2001vs} plotted in Fig 4 therein.} of the GUP-inspired Schwarzschild black hole remnant.


On the experimental front, investigation of extreme accelerations experienced by an electron during the process of radiative free neutron beta decay \cite{nico}, in the RDKII collaboration experiment \cite{RDKII:2016lpd} demonstrates proof that decay systems exist to examine specific trajectories of the electron-mirror correspondence \cite{Lynch:2022rqx}. Other such decays may correspond to the geometries we have explored here.  

Ultra-relativistically, the Analog Black Hole Evaporation via Lasers (AnaBHEL) experiment \cite{AnaBHEL:2022sri} demonstrates that the high-speed electrons \cite{Chen:2015bcg} associated with the moving mirror play an important role as a probe for spectral analysis.  These efforts could result in new physics associated with quantum vacuum radiation using an accelerated plasma \cite{Chen:2020sir}.  

\begin{acknowledgments} 

Funding comes in part from the FY2024-SGP-1-STMM Faculty Development Competitive Research Grant (FDCRGP) no.201223FD8824 at Nazarbayev University. This work is, in addition, supported by ROC (Taiwan) Ministry of Science and Technology (MOST), Grant no.112-2112-M-002 -013, National Center for Theoretical Sciences (NCTS), and Leung Center for Cosmology and Particle Astrophysics (LeCosPA) of National Taiwan University.  

\end{acknowledgments}


\bibliography{main} 
\end{document}